\expandafter \def \csname CHAPLABELintro\endcsname {1}
\expandafter \def \csname EQLABELquintic\endcsname {1.1?}
\expandafter \def \csname EQLABELUpert\endcsname {1.2?}
\expandafter \def \csname EQLABELformzeta\endcsname {1.3?}
\expandafter \def \csname CHAPLABELDwork\endcsname {2}
\expandafter \def \csname EQLABELFdef\endcsname {2.1?}
\expandafter \def \csname EQLABELFseries\endcsname {2.2?}
\expandafter \def \csname EQLABELfieldrel\endcsname {2.3?}
\expandafter \def \csname EQLABELnustarsum\endcsname {2.4?}
\expandafter \def \csname FIGLABELconeK\endcsname {2.1?}
\expandafter \def \csname EQLABELchardef\endcsname {2.5?}
\expandafter \def \csname EQLABELnustarG\endcsname {2.6?}
\expandafter \def \csname EQLABELnustartrace\endcsname {2.7?}
\expandafter \def \csname EQLABELDworkZ\endcsname {2.8?}
\expandafter \def \csname CHAPLABELDwork\endcsname {3}
\expandafter \def \csname EQLABELaltprod\endcsname {3.1?}
\expandafter \def \csname EQLABELinnerprod\endcsname {3.2?}
\expandafter \def \csname CHAPLABELcalc\endcsname {4}
\expandafter \def \csname EQLABELtildeUone\endcsname {4.1?}
\expandafter \def \csname EQLABELnicond\endcsname {4.2?}
\expandafter \def \csname EQLABELcoho\endcsname {4.3?}
\expandafter \def \csname EQLABELDrels\endcsname {4.4?}
\expandafter \def \csname EQLABELcohozero\endcsname {4.5?}
\expandafter \def \csname EQLABELQpowers\endcsname {4.6?}
\expandafter \def \csname EQLABELQleqn\endcsname {4.7?}
\expandafter \def \csname EQLABELQleqntwo\endcsname {4.8?}
\expandafter \def \csname EQLABELUphi\endcsname {4.9?}
\expandafter \def \csname EQLABELUtilde\endcsname {4.10?}
\expandafter \def \csname EQLABELUzerov\endcsname {4.11?}
\expandafter \def \csname EQLABELXveqn\endcsname {4.12?}
\expandafter \def \csname EQLABELXveqntwo\endcsname {4.13?}
\expandafter \def \csname CHAPLABELopen\endcsname {5}
\magnification=1200

\font\eightrm=cmr8 at 8pt
\font\fourteenrm=cmr12 at 14pt
\font\seventeenrm=cmr17 at 17pt
\font\twentyonerm=cmr17 at 21pt

\font\ss=cmss10

\font\csc=cmcsc10

\font\twelvecal=cmsy10 at 12pt

\font\twelvemath=cmmi12

\font\fourteenbold=cmbx12 at 14pt
\font\seventeenbold=cmbx7 at 17pt

\font\fively=lasy5
\font\sevenly=lasy7
\font\tenly=lasy10

\textfont10=\tenly
\scriptfont10=\sevenly    
\scriptscriptfont10=\fively
\parskip=10pt
\parindent=20pt
\def\today{\ifcase\month\or January\or February\or March\or April\or May\or June
       \or July\or August\or September\or October\or November\or December\fi
       \space\number\day, \number\year}

\def\title#1{\footline={\ifnum\pageno<2\hfil
       \else\hss\tenrm\folio\hss\fi}\vskip1truein\centerline{{#1}   
       \footnote{\raise1ex\hbox{*}}{\eightrm Supported in part
       by the Robert A. Welch Foundation and N.S.F. Grants 
       PHY-880637 and\break PHY-8605978.}}}

\def\Z{\hfill\break}
\def\newpage{\vfill\eject}
\def\abstract#1{\centerline{\bf ABSTRACT}\vskip.2truein{\narrower\noindent#1
       \smallskip}}

\def\runninghead#1#2{\voffset=2\baselineskip\nopagenumbers
       \headline={\ifodd\pageno\rightheadline\else \leftheadline\fi}
       \def\rightheadline{{\sl#1}\hfill{\rm\folio}}
       \def\leftheadline{{\rm\folio}\hfill{\sl#2}}}
\def\SS{\mathhexbox278}

\newcount\footnoteno
\def\Footnote#1{\advance\footnoteno by 1
                \let\SF=\empty 
                \ifhmode\edef\SF{\spacefactor=\the\spacefactor}\/\fi
                $^{\the\footnoteno}$\ignorespaces
                \SF\vfootnote{$^{\the\footnoteno}$}{#1}}

\def\figbox#1#2#3{\vbox{\vskip15pt
                   \vbox{\hrule
                    \hbox{\vrule
                     \vbox{\vskip12truept\centerline #1 \vskip6truept
                          {\hskip.4truein\vbox{\hsize=5truein\noindent
                          {\bf Figure\hskip5truept#2:}\hskip5truept#3}}
                     \vskip18truept}
                    \vrule}
                   \hrule}}}
\def\place#1#2#3{\vbox to0pt{\kern-\parskip\kern-7pt
                             \kern-#2truein\hbox{\kern#1truein #3}
                             \vss}\nointerlineskip}
\def\figurecaption#1#2{\kern.75truein\vbox{\hsize=5truein\noindent{\bf Figure
    \figlabel{#1}:} #2}}
\def\tablecaption#1#2{\kern.75truein\lower12truept\hbox{\vbox{\hsize=5truein
    \noindent{\bf Table\hskip5truept\tablabel{#1}:} #2}}}
\def\boxed#1{\lower3pt\hbox{
                       \vbox{\hrule\hbox{\vrule
                         \vbox{\kern2pt\hbox{\kern3pt#1\kern3pt}\kern3pt}\vrule}
                         \hrule}}}


\def\a{\alpha}
\def\b{\beta}
\def\g{\gamma}\def\G{\Gamma}
\def\d{\delta}\def\D{\Delta}
\def\e{\epsilon}
\def\z{\zeta}

\def\Th{\Theta}

\def\l{\lambda}\def\L{\Lambda}
\def\m{\mu}
\def\n{\nu}
\def\x{\xi}

\def\p{\pi}
\def\r{\rho}

\def\ph{\phi}\def\Ph{\Phi}\def\vph{\varphi}
\def\ch{\chi}
\def\ps{\psi}\def\Ps{\Psi}


\def\ca#1{\relax\ifmmode {{\cal #1}}\else $\cal #1$\fi}

\def\calb{{\cal B}}

\def\calm{{\cal M}}


\def\inbar{\vrule height1.5ex width.4pt depth0pt}
\def\IB{\relax{\rm I\kern-.18em B}}
\def\IC{\relax\hbox{\kern.25em$\inbar\kern-.3em{\rm C}$}}
\def\ID{\relax{\rm I\kern-.18em D}}
\def\IE{\relax{\rm I\kern-.18em E}}
\def\IF{\relax{\rm I\kern-.18em F}}
\def\IG{\relax\hbox{\kern.25em$\inbar\kern-.3em{\rm G}$}}
\def\IH{\relax{\rm I\kern-.18em H}}
\def\II{\relax{\rm I\kern-.18em I}}
\def\IK{\relax{\rm I\kern-.18em K}}
\def\IL{\relax{\rm I\kern-.18em L}}
\def\IM{\relax{\rm I\kern-.18em M}}
\def\IN{\relax{\rm I\kern-.18em N}}
\def\IO{\relax\hbox{\kern.25em$\inbar\kern-.3em{\rm O}$}}
\def\IP{\relax{\rm I\kern-.18em P}}
\def\IQ{\relax\hbox{\kern.25em$\inbar\kern-.3em{\rm Q}$}}
\def\IR{\relax{\rm I\kern-.18em R}}
\def\IZ{\relax\ifmmode\hbox{\ss Z\kern-.4em Z}\else{\ss Z\kern-.4em Z}\fi}
\def\IGa{\relax{\rm I}\kern-.18em\Gamma}
\def\IPi{\relax{\rm I}\kern-.18em\Pi}
\def\ITh{\relax\hbox{\kern.25em$\inbar\kern-.3em\Theta$}}
\def\IOm{\relax\thinspace\inbar\kern1.95pt\inbar\kern-5.525pt\Omega}


\def\noblackboxes{\overfullrule=0pt}
\def\define{\buildrel\rm def\over =}

\def\cy{Calabi--Yau} 

\def\cys{Calabi--Yau manifolds}

\def\H#1#2{\relax\ifmmode {H^{#1#2}}\else $H^{#1 #2}$\fi}
\def\M{\relax\ifmmode{\calm}\else $\calm$\fi}

\def\Bigcheck{\lower3.8pt\hbox{\smash{\hbox{{\twentyonerm \v{}}}}}}
\def\bigboldcheck{\smash{\hbox{{\seventeenbold\v{}}}}}

\def\Bighat{\lower3.8pt\hbox{\smash{\hbox{{\twentyonerm \^{}}}}}}

\def\Msharp{\relax\ifmmode{\calm^\sharp}\else $\smash{\calm^\sharp}$\fi}
\def\Mflat{\relax\ifmmode{\calm^\flat}\else $\smash{\calm^\flat}$\fi}
\def\preMcheck{\kern2pt\hbox{\Bigcheck\kern-12pt{$\cal M$}}}
\def\Mcheck{\relax\ifmmode\preMcheck\else $\preMcheck$\fi}
\def\preMhat{\kern2pt\hbox{\Bighat\kern-12pt{$\cal M$}}}
\def\Mhat{\relax\ifmmode\preMhat\else $\preMhat$\fi}

\def\Bsharp{\relax\ifmmode{\calb^\sharp}\else $\calb^\sharp$\fi}
\def\Bflat{\relax\ifmmode{\calb^\flat}\else $\calb^\flat$ \fi}
\def\preBcheck{\hbox{\Bigcheck\kern-9pt{$\cal B$}}}
\def\Bcheck{\relax\ifmmode\preBcheck\else $\preBcheck$\fi}
\def\preBhat{\hbox{\Bighat\kern-9pt{$\cal B$}}}
\def\Bhat{\relax\ifmmode\preBhat\else $\preBhat$\fi}

\def\figBcheck{\kern3pt\hbox{\raise1pt\hbox{\bigboldcheck}\kern-11pt
    {\twelvecal B}}}
\def\figBsharp{{\twelvecal B}\raise5pt\hbox{$\twelvemath\sharp$}}
\def\figBflat{{\twelvecal B}\raise5pt\hbox{$\twelvemath\flat$}}

\def\gcheck{\hbox{\lower2.5pt\hbox{\Bigcheck}\kern-8pt$\g$}}
\def\lhat{\hbox{\raise.5pt\hbox{\Bighat}\kern-8pt$\l$}}

\def\Fcheck{\kern2pt\hbox{\raise1pt\hbox{\Bigcheck}\kern-10pt{$\cal F$}}}
\def\Fhat{\kern2pt\hbox{\raise1pt\hbox{\Bighat}\kern-10pt{$\cal F$}}}
 
\def\cp#1{\relax\ifmmode {\IP\kern-2pt{}_{#1}}\else $\IP\kern-2pt{}_{#1}$\fi}
\def\h#1#2{\relax\ifmmode {b_{#1#2}}\else $b_{#1#2}$\fi}
\def\Z{\hfill\break}

\def\tr{{\rm tr}}
\def\frac#1#2{{#1\over #2}}

\def\cone{\relax\thinspace\hbox{$<\kern-.8em{)}$}}
\mathchardef\mho"0A30

\def\-{\hphantom{-}}



\def\picture #1 by #2 (#3){\vbox to #2{\hrule width #1 height 0pt depth 0pt
                                       \vfill\special{picture #3}}}
\def\scaledpicture #1 by #2 (#3 scaled #4){{\dimen0=#1 \dimen1=#2
           \divide\dimen0 by 1000 \multiply\dimen0 by #4
            \divide\dimen1 by 1000 \multiply\dimen1 by #4
            \picture \dimen0 by \dimen1 (#3 scaled #4)}}
\def\illustration #1 by #2 (#3){\vbox to #2{\hrule width #1 height 0pt depth 0pt
                                       \vfill\special{illustration #3}}}
\def\scaledillustration #1 by #2 (#3 scaled #4){{\dimen0=#1 \dimen1=#2
           \divide\dimen0 by 1000 \multiply\dimen0 by #4
            \divide\dimen1 by 1000 \multiply\dimen1 by #4
            \illustration \dimen0 by \dimen1 (#3 scaled #4)}}


\def\delaOssa{\nobreak\vskip1truein\hbox to\hsize
       {\hskip 4truein Xenia de la Ossa\hfill}}

\def\hoy{\number\day\space de \ifcase\month\or enero\or febrero\or marzo\or
       abril\or mayo\or junio\or julio\or agosto\or septiembre\or octubre\or
       noviembre\or diciembre\fi\space de \number\year}


\def\cropen#1{\crcr\noalign{\vskip #1}}

\newif\ifproofmode
\proofmodefalse

\newif\ifforwardreference
\forwardreferencefalse

\newif\ifchapternumbers
\chapternumbersfalse

\newif\ifcontinuousnumbering
\continuousnumberingfalse

\newif\iffigurechapternumbers
\figurechapternumbersfalse

\newif\ifcontinuousfigurenumbering
\continuousfigurenumberingfalse

\newif\iftablechapternumbers
\tablechapternumbersfalse

\newif\ifcontinuoustablenumbering
\continuoustablenumberingfalse

\font\eqsixrm=cmr6

\def\marginstyle{\eqsixrm}

\newtoks\chapletter
\newcount\chapno
\newcount\eqlabelno
\newcount\figureno
\newcount\tableno

\chapno=0
\eqlabelno=0
\figureno=0
\tableno=0

\def\chapfolio{\ifnum\chapno>0 \the\chapno\else\the\chapletter\fi}

\def\bumpchapno{\ifnum\chapno>-1 \global\advance\chapno by 1
\else\global\advance\chapno by -1 \setletter\chapno\fi
\ifcontinuousnumbering\else\global\eqlabelno=0 \fi
\ifcontinuousfigurenumbering\else\global\figureno=0 \fi
\ifcontinuoustablenumbering\else\global\tableno=0 \fi}

\def\setletter#1{\ifcase-#1{}\or{}%
\or\global\chapletter={A}%
\or\global\chapletter={B}%
\or\global\chapletter={C}%
\or\global\chapletter={D}%
\or\global\chapletter={E}%
\or\global\chapletter={F}%
\or\global\chapletter={G}%
\or\global\chapletter={H}%
\or\global\chapletter={I}%
\or\global\chapletter={J}%
\or\global\chapletter={K}%
\or\global\chapletter={L}%
\or\global\chapletter={M}%
\or\global\chapletter={N}%
\or\global\chapletter={O}%
\or\global\chapletter={P}%
\or\global\chapletter={Q}%
\or\global\chapletter={R}%
\or\global\chapletter={S}%
\or\global\chapletter={T}%
\or\global\chapletter={U}%
\or\global\chapletter={V}%
\or\global\chapletter={W}%
\or\global\chapletter={X}%
\or\global\chapletter={Y}%
\or\global\chapletter={Z}\fi}

\def\tempsetletter#1{\ifcase-#1{}\or{}%
\or\global\chapletter={A}%
\or\global\chapletter={B}%
\or\global\chapletter={C}%
\or\global\chapletter={D}%
\or\global\chapletter={E}%
\or\global\chapletter={F}%
\or\global\chapletter={G}%
\or\global\chapletter={H}%
\or\global\chapletter={I}%
\or\global\chapletter={J}%
\or\global\chapletter={K}%
\or\global\chapletter={L}%
\or\global\chapletter={M}%
\or\global\chapletter={N}%
\or\global\chapletter={O}%
\or\global\chapletter={P}%
\or\global\chapletter={Q}%
\or\global\chapletter={R}%
\or\global\chapletter={S}%
\or\global\chapletter={T}%
\or\global\chapletter={U}%
\or\global\chapletter={V}%
\or\global\chapletter={W}%
\or\global\chapletter={X}%
\or\global\chapletter={Y}%
\or\global\chapletter={Z}\fi}

\def\chapshow#1{\ifnum#1>0 \relax#1%
\else{\tempsetletter{\number#1}\chapno=#1\chapfolio}\fi}

\def\chaplabel#1{\bumpchapno\ifproofmode\ifforwardreference
\immediate\write1{\noexpand\expandafter\noexpand\def
\noexpand\csname CHAPLABEL#1\endcsname{\the\chapno}}\fi\fi
\global\expandafter\edef\csname CHAPLABEL#1\endcsname
{\the\chapno}\ifproofmode\llap{\hbox{\marginstyle #1\ }}\fi\chapfolio}

\def\eqnum{\global\advance\eqlabelno by 1
\eqno(\ifchapternumbers\chapfolio.\fi\the\eqlabelno)}

\def\eqlabel#1{\global\advance\eqlabelno by 1 \ifproofmode\ifforwardreference
\immediate\write1{\noexpand\expandafter\noexpand\def
\noexpand\csname EQLABEL#1\endcsname{\the\chapno.\the\eqlabelno?}}\fi\fi
\global\expandafter\edef\csname EQLABEL#1\endcsname
{\the\chapno.\the\eqlabelno?}\eqno(\ifchapternumbers\chapfolio.\fi
\the\eqlabelno)\ifproofmode\rlap{\hbox{\marginstyle #1}}\fi}

\def\eqalignnum{\global\advance\eqlabelno by 1
&(\ifchapternumbers\chapfolio.\fi\the\eqlabelno)}

\def\eqalignlabel#1{\global\advance\eqlabelno by 1 \ifproofmode 
\ifforwardreference\immediate\write1{\noexpand\expandafter\noexpand\def
\noexpand\csname EQLABEL#1\endcsname{\the\chapno.\the\eqlabelno?}}\fi\fi
\global\expandafter\edef\csname EQLABEL#1\endcsname
{\the\chapno.\the\eqlabelno?}&(\ifchapternumbers\chapfolio.\fi
\the\eqlabelno)\ifproofmode\rlap{\hbox{\marginstyle #1}}\fi}

\def\eqref#1{\hbox{(\ifundefined{EQLABEL#1}***)\ifproofmode\ifforwardreference%
\else\write16{ ***Undefined Equation Reference #1*** }\fi
\else\write16{ ***Undefined Equation Reference #1*** }\fi
\else\edef\LABxx{\getlabel{EQLABEL#1}}%
\def\LAByy{\expandafter\stripchap\LABxx}\ifchapternumbers%
\chapshow{\LAByy}.\expandafter\stripeq\LABxx%
\else\ifnum\number\LAByy=\chapno\relax\expandafter\stripeq\LABxx%
\else\chapshow{\LAByy}.\expandafter\stripeq\LABxx\fi\fi)\fi}%
\ifproofmode\write2{Equation #1}\fi}

\def\fignum{\global\advance\figureno by 1
\relax\iffigurechapternumbers\chapfolio.\fi\the\figureno}

\def\figlabel#1{\global\advance\figureno by 1
\relax\ifproofmode\ifforwardreference
\immediate\write1{\noexpand\expandafter\noexpand\def
\noexpand\csname FIGLABEL#1\endcsname{\the\chapno.\the\figureno?}}\fi\fi
\global\expandafter\edef\csname FIGLABEL#1\endcsname
{\the\chapno.\the\figureno?}\iffigurechapternumbers\chapfolio.\fi
\ifproofmode\llap{\hbox{\marginstyle#1
\kern1.2truein}}\relax\fi\the\figureno}

\def\figref#1{\hbox{\ifundefined{FIGLABEL#1}!!!!\ifproofmode\ifforwardreference%
\else\write16{ ***Undefined Figure Reference #1*** }\fi
\else\write16{ ***Undefined Figure Reference #1*** }\fi
\else\edef\LABxx{\getlabel{FIGLABEL#1}}%
\def\LAByy{\expandafter\stripchap\LABxx}\iffigurechapternumbers%
\chapshow{\LAByy}.\expandafter\stripeq\LABxx%
\else\ifnum \number\LAByy=\chapno\relax\expandafter\stripeq\LABxx%
\else\chapshow{\LAByy}.\expandafter\stripeq\LABxx\fi\fi\fi}%
\ifproofmode\write2{Figure #1}\fi}

\def\tabnum{\global\advance\tableno by 1
\relax\iftablechapternumbers\chapfolio.\fi\the\tableno}

\def\tablabel#1{\global\advance\tableno by 1
\relax\ifproofmode\ifforwardreference
\immediate\write1{\noexpand\expandafter\noexpand\def
\noexpand\csname TABLABEL#1\endcsname{\the\chapno.\the\tableno?}}\fi\fi
\global\expandafter\edef\csname TABLABEL#1\endcsname
{\the\chapno.\the\tableno?}\iftablechapternumbers\chapfolio.\fi
\ifproofmode\llap{\hbox{\marginstyle#1
\kern1.2truein}}\relax\fi\the\tableno}

\def\tabref#1{\hbox{\ifundefined{TABLABEL#1}!!!!\ifproofmode\ifforwardreference%
\else\write16{ ***Undefined Table Reference #1*** }\fi
\else\write16{ ***Undefined Table Reference #1*** }\fi
\else\edef\LABtt{\getlabel{TABLABEL#1}}%
\def\LABTT{\expandafter\stripchap\LABtt}\iftablechapternumbers%
\chapshow{\LABTT}.\expandafter\stripeq\LABtt%
\else\ifnum\number\LABTT=\chapno\relax\expandafter\stripeq\LABtt%
\else\chapshow{\LABTT}.\expandafter\stripeq\LABtt\fi\fi\fi}%
\ifproofmode\write2{Table#1}\fi}

\newdimen\sectionskip     \sectionskip=20truept
\newcount\sectno
\def\section#1#2{\sectno=0 \null\vskip\sectionskip
    \centerline{\chaplabel{#1}.~~{\bf#2}}\nobreak\vskip.2truein
    \noindent\ignorespaces}

\def\advancesectno{\global\advance\sectno by 1}
\def\sectfolio{\number\sectno}
\def\subsection#1{\goodbreak\advancesectno\null\vskip10pt
                  \noindent\chapfolio.~\sectfolio.~{\bf #1}
                  \nobreak\vskip.05truein\noindent\ignorespaces}

\def\uttg#1{\null\vskip.1truein
    \ifproofmode \line{\hfill{\bf Draft}:
    UTTG--{#1}--\number\year}\line{\hfill\today}
    \else \line{\hfill UTTG--{#1}--\number\year}
    \line{\hfill\ifcase\month\or January\or February\or March\or April\or May\or June
    \or July\or August\or September\or October\or November\or December\fi
    \space\number\year}\fi}

\def\contents{\noindent
   {\bf Contents\Z}\nobreak\vskip.05truein\noindent\ignorespaces}

\def\getlabel#1{\csname#1\endcsname}
\def\ifundefined#1{\expandafter\ifx\csname#1\endcsname\relax}
\def\stripchap#1.#2?{#1}
\def\stripeq#1.#2?{#2}

\catcode`@=11 
\def\space@ver#1{\let\@sf=\empty\ifmmode#1\else\ifhmode%
\edef\@sf{\spacefactor=\the\spacefactor}\unskip${}#1$\relax\fi\fi}
\newcount\referencecount     \referencecount=0
\newif\ifreferenceopen       \newwrite\referencewrite
\newtoks\rw@toks
\def\refmark#1{\relax[#1]}
\def\refend{\refmark{\number\referencecount}}
\newcount\lastrefsbegincount \lastrefsbegincount=0
\def\refsend{\refmark{\count255=\referencecount%
\advance\count255 by -\lastrefsbegincount%
\ifcase\count255 \number\referencecount%
\or\number\lastrefsbegincount,\number\referencecount%
\else\number\lastrefsbegincount-\number\referencecount\fi}}
\def\refch@ck{\chardef\rw@write=\referencewrite
\ifreferenceopen\else\referenceopentrue
\immediate\openout\referencewrite=referenc.texauxil \fi}
%
{\catcode`\^^M=\active 
  \gdef\obeyendofline{\catcode`\^^M\active \let^^M\ }}%
%
{\catcode`\^^M=\active 
  \gdef\ignoreendofline{\catcode`\^^M=5}}
{\obeyendofline\gdef\rw@start#1{\def\t@st{#1}\ifx\t@st\blankend%
\endgroup\@sf\relax\else\ifx\t@st\bl@nkend\endgroup\@sf\relax%
\else\rw@begin#1
\backtotext
\fi\fi}}
{\obeyendofline\gdef\rw@begin#1
{\def\n@xt{#1}\rw@toks={#1}\relax%
\rw@next}}
\def\blankend{}
{\obeylines\gdef\bl@nkend{
}}
\newif\iffirstrefline  \firstreflinetrue
\def\rwr@teswitch{\ifx\n@xt\blankend\let\n@xt=\rw@begin%
\else\iffirstrefline\global\firstreflinefalse%
\immediate\write\rw@write{\noexpand\obeyendofline\the\rw@toks}%
\let\n@xt=\rw@begin%
\else\ifx\n@xt\rw@@d \def\n@xt{\immediate\write\rw@write{%
\noexpand\ignoreendofline}\endgroup\@sf}%
\else\immediate\write\rw@write{\the\rw@toks}%
\let\n@xt=\rw@begin\fi\fi\fi}
\def\rw@next{\rwr@teswitch\n@xt}
\def\rw@@d{\backtotext} \let\rw@end=\relax
\let\backtotext=\relax

\newdimen\refindent     \refindent=30pt
\def\Textindent#1{\noindent\llap{#1\enspace}\ignorespaces}
\def\refitem#1{\par\hangafter=0 \hangindent=\refindent\Textindent{#1}}
\def\REFNUM#1{\space@ver{}\refch@ck\firstreflinetrue%
\global\advance\referencecount by 1 \xdef#1{\the\referencecount}}
\def\refnum#1{\space@ver{}\refch@ck\firstreflinetrue%
\global\advance\referencecount by 1\xdef#1{\the\referencecount}\refend}

\def\REF#1{\REFNUM#1%
\immediate\write\referencewrite{%
\noexpand\refitem{#1.}}%
\begingroup\obeyendofline\rw@start}
\def\ref{\refnum\?%
\immediate\write\referencewrite{\noexpand\refitem{\?.}}%
\begingroup\obeyendofline\rw@start}
\def\Ref#1{\refnum#1%
\immediate\write\referencewrite{\noexpand\refitem{#1.}}%
\begingroup\obeyendofline\rw@start}
\def\REFS#1{\REFNUM#1\global\lastrefsbegincount=\referencecount%
\immediate\write\referencewrite{\noexpand\refitem{#1.}}%
\begingroup\obeyendofline\rw@start}

\def\REFSCON#1{\REF#1}

\def\cite#1{\refmark#1}
\catcode`@=12 
%
%
\input epsf.tex
\hsize=6.5truein
\vsize=9truein
\proofmodefalse
\baselineskip=15pt plus 1pt minus 1pt
\parskip=5pt
\parindent=0pt
\chapternumberstrue
\forwardreferencefalse
\figurechapternumberstrue
\tablechapternumberstrue
\ifproofmode
\immediate\openout2=allcrossreferfile \fi
\ifforwardreference\input labelfile
\ifproofmode\immediate\openout1=labelfile \fi\fi
\noblackboxes
\hfuzz=1pt
\vfuzz=2pt
\def\hourandminute{\count255=\time\divide\count255 by 60
\xdef\hour{\number\count255}
\multiply\count255 by -60\advance\count255 by\time
\hour:\ifnum\count255<10 0\fi\the\count255}

\def\chaplabel#1{\bumpchapno\ifproofmode\ifforwardreference
\immediate\write1{\noexpand\expandafter\noexpand\def
\noexpand\csname CHAPLABEL#1\endcsname{\the\chapno}}\fi\fi
\global\expandafter\edef\csname CHAPLABEL#1\endcsname
{\the\chapno}\ifproofmode
\llap{\hbox{\marginstyle #1\ifnum\chapno > -1\ \else
\hskip1.3truein\fi}}\fi\chapfolio}

\def\section#1#2{\sectno=0 \null\vskip\sectionskip
    \ifnum\chapno > -1 
    \leftline{\fourteenrm\chaplabel{#1}.~~\fourteenbold#2}
    \else
    \leftline{\fourteenbold Appendix\ \chaplabel{#1}: {#2}}\fi
    \nobreak\vskip.2truein
    \noindent\ignorespaces}
\def\subsection#1{\goodbreak\advancesectno\null\vskip10pt
                  \noindent{\it \chapfolio.\sectfolio.~#1}
                  \nobreak\vskip.05truein\noindent\ignorespaces}
\def\subsubsection#1{\goodbreak
                  \noindent$\underline{\hbox{#1}}$
                  \nobreak\vskip-5pt\noindent\ignorespaces}
\def\cite#1{\refmark{#1}}
\def\\{\hfill\break}
\def\cropen#1{\crcr\noalign{\vskip #1}}
\def\contents{\line{{\fourteenbold Contents}\hfill}\nobreak\vskip.05truein\noindent\ignorespaces}

\def\titlebox#1#2{\lower7pt\hbox{%
\hsize=5in\vbox{\vskip5pt\centerline{#1}\vskip5pt\centerline{#2}}}}

\font\sevenrm cmr7
\font\mathbb msbm7 at 10pt
\font\sevenmathbb msbm7
\font\frak eufm10
\font\sevenfrak eufm7

\def\Fp{\hbox{\mathbb F}_{\kern-2pt p}}
\def\Fq{\hbox{\mathbb F}_{\kern-2pt q}}

\def\Fq{\hbox{\mathbb F}_q}

\def\Fqstar{\Fq^*}

\def\goth{\hbox{\frak G}}
\def\faith{\hbox{\frak F}}
\def\death{\hbox{\frak D}}\def\sevendeath{\hbox{\sevenfrak D}}
\def\youth{\hbox{\frak U}}\def\sevenyouth{\hbox{\sevenfrak U}}
\def\aith{\hbox{\frak A}}

\def\bb#1{\hbox{\mathbb #1}}

\def\bone{{\bf 1}}

\def\notdiv{\hbox{$\not|$\kern3pt}}

\def\ord#1{\ca{O}\kern-2pt\left(#1\right)}

\def\teich{\hbox{Teich}}
\def\seventeich{\hbox{\sevenrm Teich}}

\def\preNcheck{\kern2pt\hbox{\Bigcheck\kern-10pt{$N$}}}
\def\Ncheck{\relax\ifmmode\preNcheck\else $\preNcheck$\fi}
\def\preZcheck{\kern2pt\hbox{\Bigcheck\kern-10pt{$Z$}}}
\def\Zcheck{\relax\ifmmode\preZcheck\else $\preZcheck$\fi}

\def\norm#1{{\left\| #1\right\|}}
\def\frac#1{{\left\langle #1\right\rangle}}
\def\Cp{\hbox{\mathbb C}_{\kern-2pt p}}
\def\Zp{\hbox{\mathbb Z}_{\kern-2pt p}}
\def\Qp{\hbox{\mathbb Q}_{\kern-2pt p}}

\def\frac#1{{\left\langle #1\right\rangle}}

\def\C{{\bb C}}
\def\Z{{\bb Z}}

\def\bra#1{\left\langle #1\right|}
\def\ket#1{\left| #1\right\rangle}
\def\innerprod#1#2{\left\langle #1 | #2\right\rangle}
\def\Tr{\hbox{Tr}}
\def\sdet{\hbox{sdet}}

\def\arr{\longrightarrow}
\def\darr{\downarrow}
%
\nopagenumbers\pageno=0
\null\vskip-20pt
\rightline{\eightrm hep-th/yymmnnn}\vskip-3pt
\rightline{\eightrm \today}
\vskip0.6truein
\centerline{\seventeenrm The Zeta-Function of a p-Adic Manifold,}
\vskip.2truein
\centerline{\seventeenrm Dwork Theory for Physicists}
\vskip0.5truein
\centerline{\csc Philip~Candelas and Xenia~de~la~Ossa}
%
\vskip0.25truein\bigskip
\centerline{\it Mathematical Institute}
\centerline{\it Oxford University}
\centerline{\it 24-29 St.\ Giles'}
\centerline{\it Oxford OX1 3LB, England}
\vskip1.5truein
\vfill
\vbox{
\baselineskip=14pt
\centerline{\bf Abstract}
\vskip.1truein 
\hyphenation{super-determinant}
\noindent In this article we review the observation, due originally to Dwork, that the $\z$-function of an arithmetic variety, defined originally over the field with $p$ elements, is a superdeterminant. We review this
observation in the context of the family of quintic threefolds, 
$\sum_{i=1}^5 x_i^5 - \vph \prod_{i=1}^5 x_i = 0$,
and study the $\z$-function as a function of the parameter $\vph$. Owing to cancellations, the
superdeterminant of an infinite matrix reduces to the (ordinary) determinant of a finite matrix, $U(\vph)$,
corresponding to the action of the Frobenius map on certain cohomology groups.
The $\vph$-dependence of $U(\vph)$ is given by a relation $U(\vph) = E^{-1}(\vph^p)U(0)E(\vph)$ with
$E(\vph)$ a Wronskian matrix formed from the periods of the manifold. The periods are defined by series that converge for $\norm{\vph}_p < 1$. The
values of $\vph$ that are of interest are those for which $\vph^p = \vph$ so, for nonzero $\vph$, we have 
$\norm{\vph}_p=1$. We explain how the process of p-adic analytic continuation applies to this case.
The matrix $U(\vph)$ breaks up into submatrices of rank 4 and rank 2 and we are able from this perspective to explain some of the observations that have been made previously by numerical~calculation.
}
\newpage
%
%
\vbox{\baselineskip=10pt\parindent=20pt
\contents
\vskip10pt
\item{1.~}Introduction
\vskip10pt
\item{2.~}Dwork's Evaluation of the Zeta Function
\itemitem{\it 2.1~}{\it Review of Dwork's character}
\itemitem{\it 2.2~}{\it Operators on a vector space}
\vskip10pt
\item{3.~}Dwork Cohomology
\itemitem{3.1~}{\it Exterior and covariant derivatives}
\itemitem{3.2~}{\it Overconvergent series}
\itemitem{3.3~}{\it The superdeterminant of the complex}
\vskip10pt
\item{4.~}Calculation of the determinants
\itemitem{\it 4.1~}{\it The $4{\times}4$ determinant for $\vph=0$}
\itemitem{\it 4.2~}{\it Variation of structure}
\itemitem{\it 4.3~}{\it The other monomials}
\vskip10pt
\item{5.~}{Open Problems}
\itemitem{\it 5.1~}{\it Special Geometry}
\itemitem{\it 5.2~}{\it The $\vph=\infty$ limit}
\itemitem{\it 5.3~}{\it Truncated periods vs.\ infinite series}}
\newpage
%
\headline={\ifproofmode\hfil\eightrm draft:\ \today\ \hourandminute\else\hfil\fi}
\footline={\rm\hfil\folio\hfil}
\pageno=1
\section{intro}{Introduction}
A fundamental object of study for an arithmetic variety is its $\z$-function. Consider, for example, the one parameter family of quintic threefolds, $\M_\vph$, defined by the vanishing of the polynomial
 $$
 P(x,\vph)~=~\sum_{i=1}^5 x_i^5 -\vph \prod_{i=1}^5 x_i\eqlabel{quintic}$$
which is the family of manifolds that will largely occupy us here. If $\vph$ takes values in $\Fp$, the field with $p$ elements, and the manifold is considered as a submanifold of $\Fp\bb{P}^4$ then one can compute $N_1(\vph)$ the number of solutions to \eqref{quintic}. More generally one can take $\vph\in\Fp$ and the coordinates 
$x_j\in \bb{F}_{p^m}$ and denote the number of solutions by $N_m(\vph)$. The $\z$-function is defined as a generating function for these numbers
 $$
 \z(\vph,T)~=~\exp\left\{\sum_{m=1}^\infty N_m(\vph){T^m\over m}\right\}~.$$
The form of the $\z$-function as a function of $T$ is greatly restricted by the Weil conjectures~
\Ref{\WeilConj}{A. Weil,\ ``Solutions of equations in finite fields'',\\ 
Bull.\ American Math.\ Soc.\ {\bf 55} pp497--508 1949.}, 
which have since been proved. One of these conjectures, proved by Dwork~ 
\Ref{\Rationality}{B. Dwork,\ ``On the rationality of the zeta function of an algebraic variety'',\\ 
Amer. J. Math.  {\bf 82}  pp631--648 1960.}, 
states that the $\z$-function is a rational function of $T$. The proof proceeds by showing that the 
\hbox{$\z$-function}\ is a ratio of products of determinants that of the form 
$\det\left( 1- U(\vph)T\right)$ for certain finite matrices $U(\vph)$ that are independent of $T$. 

Dwork showed~
\REFS{\DworkI}{B. Dwork,``On the zeta-function of a hypersurface'',\\
Publ. Math. I.H.\'{E}.S., {\bf 12} pp5--68 1962.} 
\REFSCON{\DworkII}{B. Dwork,``On the zeta-function of a hypersurface II'',\\
Ann. Math. {\bf 80} pp227--299 1964.}
\refsend\
that the $\z$-function is a superdeterminant (though this was not stated in this language) of a matrix that expresses the action of the Frobenius map on a differential complex associated to the manifold. The Frobenius map is important in what follows so we pause to review it here. 

Consider again the manifold defined by the equation 
$P(x,\vph)=0$ where we take the coefficients of $P$ to take values in $\Fp$ (in our case this is just the statement $\vph\in\Fp$) but we allow $x$ to take values in a larger field such as the algebraic closure of $\Fp$ or, as we shall want to do later, in $\C_p$ which is the completion of the algebraic closure of $\Qp$, the field of $p$-adic numbers. Now since $P(x,\vph)=0$ we have $P(x,\vph)^p=0$ hence
$$
P(x^p,\vph)=0~,~~~\hbox{with}~~~x^p=(x_1^p,\,x_2^p,\,x_3^p,\,x_4^p,\,x_5^p)$$
and where we have used the fact that $\vph^p=\vph$.
In this way we see that the Frobenius map
 $$
 \hbox{Fr} (x)=x^p$$
is an automorphism of $\M_\vph$. The fixed points of this map are the points for which 
\hbox{$x_j^p=x_j,~j=1,\ldots,5$} and this is precisely the condition that $x$ be defined over $\Fp$.~
\REF{\RigidityThm}{For an accessible account see, for example, \SS1.2 of:\\
E. Frenkel,``Lectures on the Langlands Program and Conformal Field Theory'',\\
hep-th/0512172.}
Thus  $N_1$ is the number of fixed points of the Frobenius map.\Footnote{The rigidity theorem
\cite{\RigidityThm} states that the Frobenius map generates the full Galois group so one can equally say that $N_1$ is the number of fixed points of the Galois action.} Thus Dwork's approach provides an analogue of the Lefschetz fixed point theorem that applies to manifolds defined over $\bb{C}$. This permits the Euler number of the fixed point set of an automorphism to be calculated as the trace of a matrix representing the action of the automorphism on the cohomology of the manifold.

We work through this analysis here in the context of the quintic threefold in a way that we hope is straightforward. The quantities that are familiar in the complex case, particularly when the manifold can be embedded in a toric variety, duly make their appearance. In particular the Newton polytope of the manifold and the cone of monomials over the Newton polyhedron play an important role as do the periods of the manifold and the Picard-Fuchs equation. An intriguing difference that we do not resolve is that, in the present calculation, the periods appear as infinite series rather than in the truncated forms of~
\Ref{\CdORVI}{P. Candelas, X. de la Ossa and F. Rodr\'{\i}guez-Villegas, ``\cy\ Manifolds over Finite Fields I'', 
hep-th/0012233.}. 

Dwork represents the (inverse of) the action of the Frobenius map by a matrix $U(\vph)$ that is given in a p-adic neighborhood of $\vph=0$ by the expression
$$
U(\vph)~=~E^{-1}(\vph^p)U(0)E(\vph)\eqlabel{Upert}$$
where $E$ is the Wronskian matrix formed from the periods of the manifold. The periods are defined by series, and the series converge on the disk $\norm{\vph}_p<1$. If the series were to converge also for the 
Teichm\"uller points for which $\vph^p=\vph$ then the $\z$-function, which is calculated in terms of determinants of the form $\det\left( 1- U(\vph)T\right)$ would not depend on~$\vph$. In fact the series diverge for $\norm{\vph}_p=1$ and the matrix $U(\vph)$ has to be defined via a process of $p$-adic analytic continuation. This leads to a matrix $U(\vph)$ which is defined on the Teichm\"uller points but which is no longer of the form \eqref{Upert}.

In~ 
\Ref{\CdORVII}{P. Candelas, X. de la Ossa and F. Rodr\'{\i}guez-Villegas, ``\cy\ Manifolds over Finite Fields II'', 
in ``Toronto 2001, Calabi-Yau Varieties and Mirror Symmetry'', pp121-157, hep-th/0402133.}\ 
the $\z$-function was calculated for the family \eqref{quintic}.  This was done numerically in virtue of the fact that the numbers of points $N_m(\vph)$ can be computed rapidly in terms of the periods of $\M_\vph$ (at least for $p$ not too large and for the first few values of $m$). This was sufficient to suggest that the $\z$-function factorizes in the form
 $$
  \z(\vph,T)~=~{R_{\bf 1}\left(\vph,T\right)\, R_A\left(\vph,p^\r T^\r\right)^{20\over\r}
  \, R_B\left(\vph,p^\r T^\r\right)^{30\over\r}\over (1-T)(1-pT)(1-p^2 T)(1-p^3 T)}~, \eqlabel{formzeta}
 $$
the interesting point here is the form of the numerator since the form of the denominator is standard. In this expression $\r$ $(=1,2~\hbox{or}~4)$ is the smallest integer such that $5|p^\r -1$ and
the quantities $R_{\bf 1},\,R_A,\,R_B$ are quartics in their second argument. For example, $R_{\bf 1}$  takes the~form
 $$
 R_{\bf 1}(\vph,T)~=~1 + a(\vph)\, T + b(\vph)\, pT^2 +  a(\vph)\, p^3 T^3 + p^6 T^4
 $$
with $a(\vph )$ and $b(\vph)$ integers that depend on $\vph$. 

The organisation of this paper is as follows: in \SS2 we review the theory of the Dwork character and show how this relates to an operator on a vector space. The vector space in question is that of power series in the coordinates $x_j$. The $\z$-function for $\M_\vph$ is the superdeterminant of a matrix acting on a complex of forms of degree up to 5. The matrices are at this stage infinite matrices acting on the infinite dimensional space of power series. By defining a suitable covariant derivative, the eigenvalues of the supermatrix are seen to cancel in the superdeterminant apart from a finite number that correspond to the action of the supermatrix on finite cohomology groups. Furthermore the superdeterminant corresponding to these cohomology groups itself decomposes into a product of factors that can be identified with the quartics 
$R_{\bf 1},\,R_A,\,R_B$ that appear in ~\eqref{formzeta}.

Our interest in this formalism arises, in part, from a desire to study the arithmetic properties of the moduli spaces of \cys. We hope to return to this topic elsewhere particularly in regard to arithmetic special geometry and arithmetic properties of the attractor mechanism. 
\newpage
\REF{\DGS}{``An Introduction to G-Functions'', by B. Dwork, G. Gerotto and F. J. Sullivan,
Annals of Mathematics Studies {\bf 133}, Princeton University Press 1994.}
\REF{\Koblitz}{``p-adic Numbers, p-adic Analysis, and Zeta Functions'', Second Edition,\\ 
N.~Koblitz, Graduate Texts in Mathematics, Springer 1984.}
\section{Dwork}{Dwork's Evaluation of the Zeta Function}
\vskip-30pt 
\subsection{Review of Dwork's character}
We begin by reviewing the properties of Dwork's character 
$\Th:\,\Fp\rightarrow\bb{C}_p^*$; a full account may be found in \cite{\DGS,\Koblitz}. 
In order to define $\Th$ we define first a function $\faith$ as a power series
 $$
\faith(X)~=~\exp\bigl(\p\big(X-X^p)\bigr)~=~\sum_{n=0}^\infty c_n (\p X)^n\eqlabel{Fdef}$$
where $\p$ is a number in $\bb{C}_p$ such that $\p^{p-1}=-p$. The exponential is understood as given by the
usual power series and the resulting power series in $X$ defines $\faith$ in the first instance. By
differentiating $\faith$ we find
 $$
{1\over\p}{d\faith\over dX}~=~\Big(1+(\p X)^{p-1}\Big)\faith$$
and it is easy to show from this relation that the series for $\faith$ has the form shown and that the
coefficients satisfy the recurrence $nc_n~=~c_{n-1} + c_{n-p}$ with $c_n=0$ for $n<0$
and $c_0=1$. Now it is an essential fact that the series on the right of \eqref{Fdef} converges in
the disk $\norm{X}_p\leq 1+\e$ for some fixed positive $\e$. The exponential series $\exp(\p Y)$ converges, for p-adic $Y$, in the disk $\norm{Y}_p<1$ and the $X$-disk that ensures $\norm{X-X^p}_p<1$ is the disk 
$\norm{X}_p<1$.
We now give an improved definition for $\faith(X)$ as the sum of the series in \eqref{Fdef} valid throughout the disk where this series converges
 $$ 
\faith(X)~=~\sum_{n=0}^\infty c_n (\p X)^n~~;~\norm{X}_p\leq 1+\e~.\eqlabel{Fseries}$$
So defined, $\faith(X)$ exists and is p-adic analytic in the disk
$\norm{X}_p\leq 1+\e$. On the smaller disk $\norm{X}_p<1$ we have
$\faith(X)=\exp\bigl(\p\big(X-X^p)\bigr)$. We shall need also to evaluate $\faith(X)$ for
$\norm{X}_p=1$. For such $X$ the series converges however, as we shall see shortly, it does
not converge to
$\exp\big(\p\bigl(X-X^p)\bigr)$. 

The character $\Th:\,\Fp\rightarrow \bb{C}^*_p$, of order $p$, is defined by
 $$
\Th(x)~=~\faith\bigl(\teich(x)\bigr)~.$$
The Teichm\"uller representative $X=\teich(x)$ corresponds to the embedding of $\Fp^*$
in $\Cp^*$ as a multiplicative group. We think of $x$ as an integer in the range 
$0\leq x\leq p-1$ and then we define $\teich(x)$ as a limit
 $$
\teich(x)~=~\lim_{n\to\infty}x^{p^n}$$
which converges in the p-adic sense. Thus $\teich(x)=x+\ord{p}$ and $\teich(x)$ satisfies the equation 
 $$
\teich(x)^p=\teich(x)~.\eqlabel{fieldrel}$$ 
In virtue of our previous discussion however we are prepared for the fact that even
though this last relation holds nevertheless $\Th(x)\neq\exp(0)=1$. In fact we see that
 $$
\Th(1)~=~1 + \p + \ord{\p^2}$$
and since $\norm{\p}_p < 1$ it follows that $\Th(1)\neq 1$. If, however, $X^p=X$ then
 $$
\faith(X)^p~=~\exp(p\p X - p\p X^p)~=~1$$
since the presence of the $p$ in the exponent ensures the convergence of the series. In particular $\Th(1)$
is a $p$'th root of unity and since $\Th(x)=1+\p\,\teich(x) +\ord{\p^2}$ 
we see also that $\Th(x)=\Th(1)^{\seventeich(x)}=\Th(1)^x$. 

It follows from the definition of the Teichm\"uller representative that for $x,y\in\Fp$
 $$
\teich(x+y) ~=~ \teich(x) + \teich(y) + pZ $$
for some $Z\in \bb{Z}_p$. From this we see that $\Th$ is a nontrivial additive
character, that is
 $$
\Th(x+y)~=~\Th(x)\,\Th(y)~,$$
and moreover is of order $p$  since $\Th(px)= 1$.

Dwork also adapted this construction to give a character, of order $p$,
$\Th_r:\,\bb{F}_{p^r}\rightarrow \bb{C}_p^*$
 $$
\Th_r(x)~=~\Th(\tr(x))$$
where $\tr:\,\bb{F}_{\kern-2pt p^r}\rightarrow\Fp$ is the trace map
 $$
\tr(x)~=~x + x^p + x^{p^2} + \cdots + x^{p^{r-1}} $$
Now the limit $\teich(x)=\lim_{n\to\infty}x^{q^n}$ exists also for $x\in\Fq$ though
$\teich(x)$, for $x\neq 0$, is now a unit (has unit norm) of $\bb{C}_p$ but is not in
general\Footnote{Recall that $\Fq$ is an extension of degree $r$ over $\Fp$. That is,
$\Fq=\Fp(\a)$ with $\a$ the root of an irreducible monic polynomial of degree $r$ with coefficients in $\Fp$. Thus
for $x\in\Fq$ the Teichm\"uller representative will have the form
$\teich(x)=\sum_{k=0}^{r-1}b_k\a^k$ with coefficients in $\Qp$ (in fact in $\Zp$) but
the root $\a$ is not in $\Qp$.} in $\bb{Z}_p$. 
We have
 $$ \teich\left(\tr(x)\right)~=~\sum_{\ell=0}^{r-1}\teich^{p^\ell}\!(x) + pZ $$
with $Z$ an integer of $\bb{C}_p$.
It follows that 
 $$
\Th_r(x)~=~\Th(1)^{\sum_{\ell=0}^{r-1}\seventeich^{p^\ell}\!(x)}~.$$
Note however that we cannot, in general, write the right hand side of this relation as
the product
 $$
\prod_{\ell=0}^{r-1} \Th(1)^{\seventeich^{p^\ell}\!(x)}$$
since the $\teich^{p^\ell}(x)$ are not in general in $\bb{Z}_p$. Dwork showed nevertheless
that $\Th_r$ has the remarkable splitting property
 $$
\Th_r(x)~=~\prod_{\ell=0}^{r-1} \Th\!\left(x^{p^\ell}\right)~.$$
The characters in the product make sense through the series
\eqref{Fseries} for $\norm{X}_p\leq 1+\e$ even if $X$ is in $\Cp$ rather than $\Qp$.

To show the utility of Dwork's character let $P(X)\in \Fp[X^1,X^2,\ldots,X^n]$ be a homogeneous polynomial of
degree $d$ and let $\n_r^*$ denote the number of $x=(x^1,\ldots,x^n)\in \bb{F}_{p^r}^n$, with no component
zero, of solutions to the equation $P(x)=0$, that is 
 $$
\n_r^*~=~\#\left\{x\in(\bb{F}_{p^r}^*)^n~~\big|~~P(x)=0\right\}~.$$
Then in virtue of the relation
 $$
\sum_{y\in\hbox{\sevenmathbb F}_{p^r}}\Th_r\big(yP(x)\big)~=~\cases{p^r~,&if $P(x)=0$~,\cropen{5pt} 
                                                       0~,&otherwise~,\cr}$$
we have
 $$
p^r \n_r^*~=~(p^r-1)^n+\hskip-10pt\sum_{(y,x)\in (\hbox{\sevenmathbb F}_{p^r}^*)^{n+1}}
\hskip-10pt\Th_r\big(yP(x)\big)~,
\eqlabel{nustarsum}$$
where the first term on the right arises from splitting off the $y=0$ contribution to the~sum. 

It is convenient to set $x^0=y$ and to write
 $$
W(x)~=~x^0\,P(x^1,\ldots,x^n)~.$$
We will have need of some notation relating to polynomials and series. For monomials we employ a multi-index
notation and write $X^{\bf v}=\prod_{\a=0}^n (X^\a)^{v_\a}$. We shall refer to $\sum_{j=1}^n v_j$ as the
degree of $X^{\bf v}$ ignoring $v_0$. The set of all $X^{\bf v}$ of degree $\ell d$ with $v_0=\ell$,
$\ell=0,1,\ldots,$ define an $n$-dimensional lattice $\L$. Within this lattice the monomials $X^{\bf v}$
that arise with nonzero coefficient in $W(X)$ define a polyhedron $\D$. We shall have need also of the cone
$K\subset\L$ subtended from ${\bf v}=0$ by $\D$. For the case of $W$ a generic cubic and $n=3$ this is
illustrated in Figure~\figref{coneK}. With these conventions we can write 
$W(x)=\sum_{{\bf v}\in\D}w_{\bf v}x^{\bf v}$ and we shall understand $W(X)$ for p-adic $X$ to be given by
 $$
W(X)~=~\sum_{{\bf v}\in\D}W_{\bf v}X^{\bf v}~,~\hbox{with}~W_{\bf v}=\teich(w_{\bf v})~.$$

\midinsert
\def\coneK{\vbox{\vskip10pt\hbox{\hskip0pt\epsfxsize=6.4truein\epsfbox{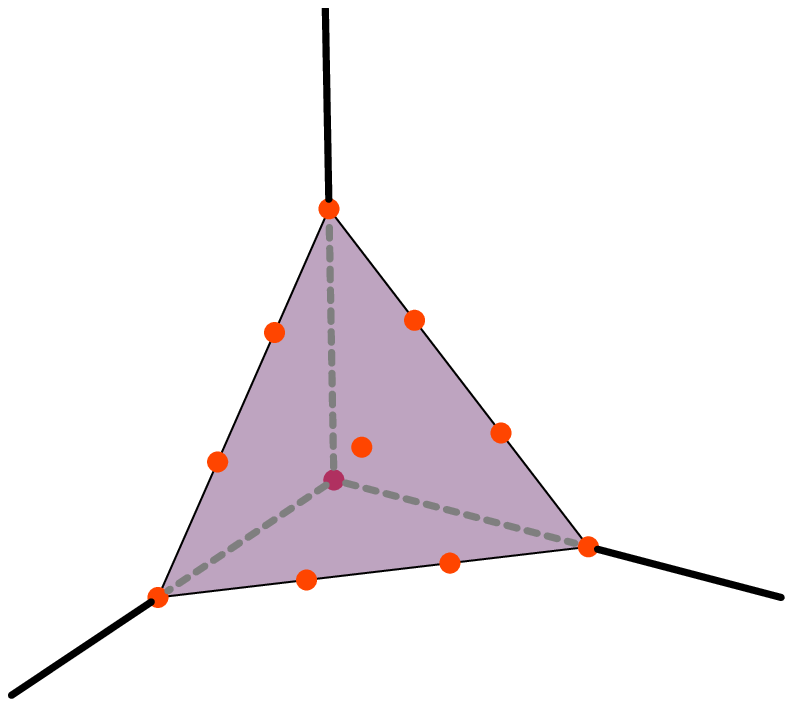}}\vskip-30pt}}
\figbox{\coneK}{\figlabel{coneK}}{The cone $K$ and polyhedron $\D$ for the generic cubic.}
\vskip20pt
\endinsert

We have explained how the character
$\Th\big(W(x)\big)$ is related to the series
$$
\faith(W(X))~=~\exp\bigl(\p\big(W(X)-W(X)^p)\bigr)~.\eqlabel{chardef}$$ 
We wish now to show also that $\Th\big(W(x)\big)=\goth(\teich(x))$ for $\goth(X)$ the function defined by the power
series
 $$
\goth(X)~=~\exp\bigl(\p\big(W(X)-W(X^p)\big)\bigr)~=~\sum_{{\bf v}\in K}G_{\bf v}X^{\bf v}~.$$
In other words we claim that the series on the right converges on the polydisk $\norm{X^\a}_p\leq 1$ though again
it does not converge, when $\norm{X}_p=1$, to $\exp\bigl(\p\big(W(X)-W(X^p)\big)\bigr)$.
To see that $\Th\big(W(x)\big)=\goth(\teich(x))$ note that
 $$\eqalign{
\Th\big(W(x)\big)~&=~\prod_{{\bf v}\in\D}\Th(w_{\bf v}x^{\bf v})\cropen{3pt}
&=~\lim_{X\to\seventeich(x)}\prod_{{\bf v}\in\D}
\exp\big[\p(W_{\bf v}X^{\bf v}-W_{\bf v}X^{p\bf v})\big]\cropen{3pt}
&=~\lim_{X\to\seventeich(x)}\exp\big[\p(W(X)-W(X^p))\big]\cropen{3pt}
&=~\lim_{X\to\seventeich(x)}\goth(X)~.\cr}$$
Similar relations apply also to the functions $\Th\left(W(x)^{p^\ell}\right)$ and in this way we see that
our expression \eqref{nustarsum} can be rewritten in the form
 $$
p^r \n_r^*~=~(p^r-1)^n + \hskip-10pt\sum_{(X^\a)^{p^r - 1}=1}
\hskip-10pt \goth\big( X\big) \goth\big( X^p\big)\ldots \goth\big( X^{p^{r-1}}\big)~.\eqlabel{nustarG}$$
\subsection{Operators on a vector space}
One of Dwork's great insights was to regard the computation of the $\z$-function as a sequence of operations
on a vector space in a way that bears an intriguing similarity to the formalism of quantum mechanics. The
analog of a Hilbert space is here a ring $\ca{H}$ of power series of the form
$\Ph(X)=\sum_{{\bf v}\in K}\Ph_{\bf v}X^{\bf v}$ where the powers $X^{\bf v}$ that occur lie in the cone $K$ and
the coefficients $\Ph_{\bf v}$ are required to decrease such that 
$\norm{\Ph_{\bf v}}_p < \norm{\p}^{\deg{\bf v}}_p$ as $\deg{\bf v}\to\infty$. 

A basis of states is provided by the monomials
$\left\{X^{\bf v}\right\}_{{\bf v}\in K}$ which, following the usage of quantum mechanics, we can think of
abstractly as states $\ket{\bf v}$. There is no notion of hermitian conjugation however we may
nevertheless define a dual basis $\bra{\bf v}$ by requiring
 $$
\langle{\bf u}|{\bf v}\rangle~=~\cases{1~,&if ${\bf u}={\bf v}$\cr
                                       0~,&otherwise~.\cr}$$
In $\ca{H}$, series $\Ps(X)$ are both states and operators since a series $\Ps(X)$ maps a state $\Ph(X)$ to the
state $\Ps(X)\Ph(X)$.

Dwork made important use of an operator $\aith_q:\ca{H}\to\ca{H}$, that seems to have been first introduced by Adkin, and which may be regarded as an inverse to the Frobenius map
 $$
\aith_q(X^{\bf v})~=~\cases{X^{{\bf v}/q}~,&if $q|{\bf v}$\cropen{3pt}
  				         0~,&otherwise~.\cr}$$
Equivalently the action of $\aith_q$ on a series $\Ph(X)=\sum_{{\bf v}\in K}\Ph_{\bf v}X^{\bf v}$ can be written
 $$
\aith_q\Ph(X)=\sum_{{\bf v}\in K}\Ph_{q{\bf v}}X^{\bf v}$$ 
Dwork uses the notation $\psi_q$ for this map and this notation is common in the literature.  This however
would be confusing in an account that attempts to draw analogies with quantum mechanics owing to the
established use of $\psi$ in the latter context to denote state~vectors.

It is straightforward to establish the operator identities:
 $$\eqalign{
1)\qquad &(\aith_q)^r~=~\aith_{q^r}\cropen{3pt}
2)\qquad &\aith_q \Ph(X^q)~=~\Ph(X) \aith_q\cropen{3pt}
3)\qquad &\aith_{q^r} \Ph(X) \Ph(X^q)\ldots \Ph(X^{q^{r-1}})~=~(\aith_q \Ph(X))^r~.\cr}$$
We shall have need also of the matrix elements of operators of the form $\aith_q \Ph(X)$. These can be written in terms of the coefficients of $\Ph$
 $$
\bra{\bf u}\aith_q \Ph(X)\ket{\bf v}~=~\Ph_{q{\bf u}-{\bf v}}$$
where it is understood that $\Ph_{\bf w}=0$ for ${\bf w}\not\in K$.

Consider now the effect of summing $\Ph(X)$ over the images of $(\Fqstar)^{n+1}$. We will see that this is
related to the trace of the matrix $\aith_q \Ph(X)$
 $$
\sum_{(X^\a)^{q-1}=1} \Ph(X)~=~(q-1)^{n+1}\sum_{{\bf v}\in K} \Ph_{(q-1){\bf v}}
~=~(q-1)^{n+1}\,\Tr\bigl(\aith_q \Ph(X)\bigr)~.$$
We wish to apply this identity to the sum that arises in \eqref{nustarG} in relation to the calculation of
$\nu^*$. In virtue of the identities above we find
 $$\eqalign{
\sum_{(X^\a)^{p^r - 1}=1}
\hskip-10pt \goth\big( X\big) \goth\big( X^p\big)\ldots \goth\big( X^{p^{r-1}}\big)~
&=~(q-1)^{n+1}\,
\Tr\left( \aith_{p^r} \goth\big( X\big) \goth\big( X^p\big)\ldots \goth\big( X^{p^{r-1}}\big)\right)
\cropen{3pt} 
&=~(q-1)^{n+1}\,\Tr\bigg( \Bigl(\aith _p   \goth(X)\Bigr)^r \bigg)~.\cr} $$
We may also write
 $$
\youth  ~=~\aith _p  \goth(X)~=~\hbox{e}^{-\p W(X)}\aith _p  \, \hbox{e}^{\p W(X)}~.$$
This last expression is useful for purposes of manipulation however it must be borne in mind that, while
$\goth(X)\in \ca{H}$, the series $\hbox{e}^{\pm\p W(X)}$ are not in $\ca{H}$. Thus it is not the case that
$\Tr(\youth  )$ is $\Tr\left( \aith _p  \right)$, for example.
 
Using our identities in \eqref{nustarG} we have
 $$
\nu_r^*~=~{(p^r-1)^n\over p^r} + {(p^r-1)^{n+1}\over p^r}\,\Tr\big(\youth^r\big)~.
\eqlabel{nustartrace}$$
We wish to use this expression to compute a function $Z^*(T)$ that is closely allied to the $\z$-function
 $$
Z^*(T)~=~\exp\left\{\sum_{r=1}^\infty {\n_r^*\, T^r\over r}\right\}~.$$
In order to do this we note that 
 $$
\exp\left\{\sum_{r=1}^\infty \Tr\big(\youth  ^r\big){T^r\over r}\right\}~=~
\exp\left\{-\Tr\log\left( 1 - \youth\, T\right)\right\}~=~
\det\left( 1 - \youth\, T\right)^{-1}~.$$
If we now expand the powers of $(p^r-1)$ in \eqref{nustartrace} by the binomial theorem we find
 $$
Z^*(T)~=~\prod_{j=0}^n\left( 1-p^{j-1}T\right)^{(-1)^{(n-j+1)}\,{n\choose j}}\,
\prod_{k=0}^{n+1}\det\left( 1- p^{k-1}T\,\youth  \right)^{(-1)^{(n-k)\,}{n+1\choose k}}~.\eqlabel{DworkZ}$$
At this stage the determinants that appear in this expression are determinants of infinite matrices however
Dwork has proved that $Z^*$ is a rational function of $T$. This is a consequence of the fact that, owing to
the alternating sign of the exponents in the last equation, determinants appear both in the numerator and
denominator and there is cancellation between all but a finite number of eigenvalues (note that the matrices
are all diagonal if $\youth  $ is). We are interested in how this cancellation comes about. This is a story
about cohomology to which we now turn.
\newpage
\section{Dwork}{Dwork Cohomology}
\vskip-20pt
\subsection{Exterior and covariant derivatives}
Following Monsky and Washnitzer~ 
\Ref{\MWcoho}{P. Monsky and G. Washnitzer, ``Formal cohomology. I. '',\\
Ann. of Math. (2)  {\bf 88}  pp181--217 1968.}, 
we define an analog of an exterior derivative that acts on power series. The
`differential forms' require for their definition formal symbols $dX^\a$. We start by defining logarithmic
derivatives and differentials 
 $$
D_\a~=~X^\a{\partial\over\partial X^\a}~,~~~\x^\a~=~{dX^\a\over X^\a}~.$$
In order to commute with $\youth  =\hbox{e}^{-\p W(X)}\,\aith _p  \, \hbox{e}^{\p W(X)}$ we take our exterior derivative to be
 $$
\death~=\hbox{e}^{-\p W(X)}dX^\a{\partial\over\partial X^\a}\hbox{e}^{\p W(X)}~=~\hbox{e}^{-\p W(X)}\x^a D_\a \hbox{e}^{\p W(X)}$$
where here and in the following a summation convention applies to repeated indices.
Spaces, $\ca{H}_k$, of $k$-forms are defined to be
 $$
\ca{H}_k~=~\left\{\Ph_{\a_1\a_2\cdots\a_k}(X)\,\x^{\a_1}\x^{\a_2}\ldots\x^{\a_k}~|~
\Ph_{\a_1\a_2\cdots\a_k}(X)=\Ph_{[\a_1\a_2\cdots\a_k]}(X)\in\ca{H}\right\} $$
and we obtain in this way a chain complex
 $$
\matrix{
0&\arr&\ca{H}_0&\arr&\ca{H}_1&\arr&\cdots&\arr&\ca{H}_n&\arr&\ca{H}_{n+1}&\arr&0\cropen{3pt}
&&\hfil\darr\hfil&&\hfil\darr\hfil&&&&\hfil\darr\hfil&&\hfil\darr\hfil&&&\cropen{3pt}
0&\arr&\ca{H}_0&\arr&\ca{H}_1&\arr&\cdots&\arr&\ca{H}_n&\arr&\ca{H}_{n+1}&\arr&0\cr}$$
with the horizontal maps being $\death $ and the vertical maps $\youth  $. Note that each $\ca{H}_k$ is,
abstractly, a tensor power of $\ca{H}_0$
 $$
\ca{H}_k~=~\ca{H}_0^{\otimes{n+1\choose k}}~.$$

Owing to the fact that $\x^\a=d\log(X^\a)$ and the effect of $\aith _p  $ is, in effect, to replace $X^\a$ by
$(X^\a)^{1/p}$ we define
 $$
\aith _p  (\x^\a)~=~{1\over p}\,\x^\a~~~\hbox{hence also}~~~
\youth(\x^\a)~=~{1\over p}\,\x^\a~.$$ 
With this convention it is easy to check that
$\death$ and $\youth$ commute. It is sufficient to check this for a form 
$\hbox{e}^{-\p W}X^{\bf v}\,\x^{\b_1}\ldots\x^{\b_k}$
 $$
\matrix{
\hbox{e}^{-\p W}X^{\bf v}\,\x^{\b_1}\ldots\x^{\b_k}&\buildrel\sevendeath\over\arr&
\hbox{e}^{-\p W}X^{\bf v}\,v_\a\x^\a\x^{\b_1}\ldots\x^{\b_k}\cropen{5pt}
\hfil\sevenyouth\darr~~\hfil&&\hfil\sevenyouth\darr~~\hfil\cropen{5pt}
{1\over p^k}\,\hbox{e}^{-\p W}\aith _p  \left(X^{\bf v}\right)\,\x^{\b_1}\ldots\x^{\b_k}&
\buildrel\sevendeath\over\arr&
{1\over p^{k+1}}\,\hbox{e}^{-\p W}\aith _p  \left(X^{\bf v}\right)\,v_\a\x^\a\x^{\b_1}\ldots\x^{\b_k}~.\cr}$$
Having been defined so as to commute with $\youth$ the operator $\death$ should be regarded as a covariant derivative. This being so a comment is in order regarding the manner in which $\death$ respects the Leibnitz rule.
If we write $\x^\a D_\a = D$ then we have
$$
\death(\ph\Ps)~=~\ph\,\death\Ps + (D\ph)\,\Ps
$$
which is appropriate if we regard $\Ps$ as a vector and $\ph$ as an operator. We learn that $\death$ acts differently on vectors and operators. Note however the necessary consistency property that the derivative of a vector $\Ph=\ph\Ps$ does not depend on the decomposition of the product into its factors.

\subsection{Overconvergent series}
In order to have a well behaved cohomology theory it is necessary to place convergence conditions on the series that arise in the differential forms. A first thought is that one should require the series to converge on the polydisks $\norm{X^\a}_p \leq 1$. This however is inadequate as we see already from the following standard one-dimensional example.
Consider the 1-form 
$$
\a~=~\sum_{n=0}^\infty p^n x^{p^n - 1}\,dx~=~d\left(\sum_{n=0}^\infty x^{p^n}\right)~.
$$
Owing to the fact that $\a$ is a 1-form and the space is one-dimensional, we have $d\a=0$. There is, however, no 0-form $\b$ such that $\a=d\b$,
since the series on the right does not converge for $\norm{x}_p=1$. Thus if we merely require convergence on the polydisks $\norm{X^\a}_p \leq 1$ there is a failure of the Poincar\'{e} Lemma: that, for a contractible space, a form that is closed is also exact. It was shown by Dwork in \cite{\DworkI} that this difficulty may be overcome by requiring that the series converge on polydisks of radius $1+\d$ for some fixed $\d>0$. These are referred to as {\sl overconvergent series}. 
For a certain choice of $\d$ the set of of overconvergent series is the ring
 $$
\ca{H}~=~\left\{ \sum_{{\bf v}\in K} a_{\bf v}\p^{\deg({\bf v})}\, X^{\bf v}~~\Big|~~ a_{\bf v}\to 0 ~~
\hbox{as}~~\deg({\bf v})\to\infty~~\right\}~.$$
In the cited work it is shown, moreover, that for this choice of $\ca{H}$ the only nontrivial cohomology occurs 
in~$\ca{H}_{n+1}$.

\subsection{The superdeterminant of the complex}
The superdeterminant of the operator $\left( 1-p^nT\,\youth  \right)$ of the complex
$\ca{H}_\bullet$ is
 $$
\sdet_{\ca{H}_\bullet}\left( 1-p^nT\,\youth\right)~=~
\prod_{\ell=0}^{n+1}{\det}_{\ca{H}_\ell}\left( 1-p^n T\,\youth\right)^{(-1)^\ell}$$ 
and we wish to show now that this superdeterminant is precisely the product of determinants that arises in
$Z^*(T)$. Referring back to \eqref{DworkZ} we see that the latter expression contains the product
 $$
\left[\; \prod_{\ell=0}^{n+1}{\det}_{\ca{H}_0}
\left( 1-p^{n-\ell}T\,\youth  \right)^{(-1)^\ell\, {n+1\choose\ell}}\;\right]^{-1}
\eqlabel{altprod}$$  
where we have written $k=n+1-\ell$ for the index in the product and we note that, in
this context, the determinants refer to matrices that act on $\ca{H}_0$. 

An eigenfunction $\Ph(X)\in\ca{H}_0$ of
$\youth$, that has eigenvalue $\m$, contributes a factor
$\left(1-p^n\m T\right)$ to ${\det}_{\ca{H}_0}\left(1-p^nT\,\youth\right)$. This
eigenfunction also gives rise to ${n+1\choose \ell}$ eigenforms in each $\ca{H}_\ell$ of
the form
$\Ph(X)\,\x^{\a_1}\ldots\x^{\a_\ell}$ each satisfying
 $$
p^n\youth  \Big(\Ph(X)\,\x^{\a_1}\ldots\x^{\a_\ell}\Big)~=~
p^{n-\ell}\m\,\Ph(X)\,\x^{\a_1}\ldots\x^{\a_\ell}~.$$
Thus $\Ph(X)$ contributes a factor 
$\prod_{\ell=0}^{n+1}\left(1-p^{n-\ell}\m T\right)^{(-1)^\ell\,{n+1\choose \ell}}$ to
the superdeterminant. Furthermore it is immediate that every eigenform 
$\Psi= \Psi_{\a_1\cdots\a_\ell}\x^{\a_1}\ldots\x^{\a_\ell}$
of $\youth  $ has coefficients $\Psi_{\a_1\cdots\a_\ell}$ in $\ca{H}_0$ that are eigenfunctions of $\youth  $.
Thus all the eigenforms are of the form $\Ph(X)\,\x^{\a_1}\ldots\x^{\a_\ell}$ and we see that the
alternating product in the square brackets in \eqref{altprod} is precisely the superdeterminant of the
complex. 
 $$
\sdet_{\ca{H}_\bullet}\left( 1-p^nT\,\youth\right)~=~
\prod_{\ell=0}^{n+1}{\det}_{\ca{H}_0}
\left( 1-p^{n-\ell}T\,\youth  \right)^{(-1)^\ell\, {n+1\choose\ell}}~.$$

If now $\Psi^{(\ell)}(X)\in\ca{H}_\ell$ is a $\ell$-form which is an eigenform of $\youth  $ with eigenvalue
$\m$ then, since $\death $ and $\youth  $ commute we have also 
 $$
\youth  \Big(\death \Psi^{(\ell)}(X)\Big)~=~\m\,\death \Psi^{(\ell)}(X)$$
and we see that if $\death \Psi^{(\ell)}(X)$ is nonzero it is a $(\ell+1)$-form with the same eigenvalue as
$\Psi^{(\ell)}(X)$. The contribution of such eigenforms cancels from the superdeterminant and in this way
we see that the superdeterminant reduces to a superdeterminant on the \hbox{$\death $-cohomology} 
of the complex.
 $$
\sdet_{\ca{H}_\bullet}\left( 1-p^nT\,\youth\right)~=~
\sdet_{\ca{H}_\bullet/\sevendeath\ca{H}_\bullet}\left(1-p^nT\,\youth\right)~.$$

Now that the states have become forms we shall define the inner product $\innerprod{\Ph}{\Ps}$ to be the
constant term in the expansion of the highest order form, that is the coefficient of $\x^0\x^1\ldots\x^n$.
Given that the only nontrivial cohomology occurs in $\ca{H}_{n+1}$ we shall principally be concerned with
computing products $\innerprod{\ph}{\Ps}$ between zero-forms \hbox{$\bra{\ph}=\ph^*(X)$} and
$(n+1)$-forms
$\ket{\Ps}=\ps(X)\,\x^0\x^1\ldots\x^n$. When this is the case we may write
 $$
\innerprod{\ph}{\Ps}~=~(\ph,\,\ps) $$
with the inner product on the right denoting the inner product defined previously for zero-forms.

Now if $\ket{\ch}={1\over n!}\ch^\a(X)\e_{\a\b_1\b_2\ldots\b_n}\x^{\b_1}\x^{\b_2}\ldots\x^{\b_n}$ is an
$n$-form then $\death \ket{\ch}=\death _\a\ch^\a\,\x^0\x^1\ldots\x^n$
and we see that, in cohomology, 
 $$
\ps(X)~\simeq~\ps(X) + \death _\a\ch^\a(X)~.$$
In order for our inner products to respect this equivalence the dual states must satisfy
 $$
(\ph,\,\death _\a\ch^\a)~=~0\eqlabel{innerprod}$$
for all $\ch^\a$.

Let $\g_{-}$ denote the operator that projects onto the space of dual states
 $$
\g_{-} X^{\bf v}~=~\cases{X^{\bf v}~~&~~if $-{\bf v}\in K~,~~\deg({\bf v})\neq 0,$\cropen{3pt}
0~~&~~otherwise\cr}$$ and let 
 $$
\death ^*_\a~=~-\g_{-}\, \hbox{e}^{\p W(X)}D_\a\, \hbox{e}^{-\p W(X)}$$
then a little thought shows that 
 $$
(\ph,\,\death _\a\ps)~=~(\death^*_\a\ph,\,\ps)$$
for all $\ph$ and $\ps$. Thus \eqref{innerprod} requires the condition $\death ^*_\a\ph^*(X)=0$ be imposed on the allowed dual states.
\newpage
\section{calc}{Calculation of the determinants}
We anticipate from \cite{\CdORVI} and \cite{\CdORVII} that the monomials of degree $5\ell$, $0\leq\ell\leq 5$, will play a special role. Among these there are 126 monomials of degree 5. The fifth powers, $x_i^5$, and the twenty monomials whose exponents are permutations of $(2,1,1,1,0)$ turn out to be trivial in cohomology. In the case of the monomials which are permutations of $(2,1,1,1,0)$ these are related to monomials that are permutations of $(4,1,0,0,0)$ and will not need to be counted separately. We are left with 101 quintic monomials. The monomial
$$
Q~\define~x_0x_1x_2x_3x_4x_5
$$
plays a special role, and there are 100 others that we denote by $x^{\bf v}$.  We anticipate also that the
determinants which are in principle of size 
$204{\times}204$ break up into a $4{\times}4$ block, corresponding to a basis $\{Q,Q^2,Q^3,Q^4\}$, and 100 blocks of size $2{\times}2$ corresponding to bases $\{x^{\bf v},Qx^{\bf v}\}$, one for each of the
quintic monomials~$x^{\bf v}$. We begin by considering the case $\vph=0$.
\subsection{The $4{\times}4$ determinant for $\vph=0$}
When $\vph=0$ the operator whose determinant we wish to compute is 
 $$
\youth(0)~=~\aith _p  \, \prod_{i=1}^5 \faith(X_0X_i^5)~.$$
Consider the effect of this operator on a basis $\ket{\ell}=(-\p Q)^\ell$.
 $$
\youth(0)\!\ket{\ell}~=~(-1)^\ell\, \aith _p  \,\sum_{n_i}\left(\prod_i c_{n_i}\right)\,
\p^{\sum_i n_i + \ell}
\, X_0^{\sum_i n_i + \ell}\prod_i X_i^{5n_i+\ell}~.\eqlabel{tildeUone}$$
The terms in the sum that survive the effect of $\aith _p  $ are those for which 
 $$
5n_i+\ell~\equiv~0~~\bmod p~~,~~i=1,\dots,5~.\eqlabel{nicond}$$ 
Note that these conditions imply $\sum_i n_i + \ell \equiv 0 \bmod p$ on summation over $i$.
To proceed it is useful to define integers $1<a<p-1$ and $1\leq b\leq 4$ such that $a$ is the smallest positive integer that represents $-1/5$ in $\Fp$ and $b$ satisfies
 $$
5a+1~=~bp~.$$
Then \eqref{nicond} is solved by writing
 $$
n_i~=~n(\ell) + k_i p$$
where
$$ 
n(\ell)~=~a\ell - p\left[{a\ell\over p}\right]$$
is the smallest positive representative of $a\ell$ mod $p$. Since $5n(\ell)+\ell\equiv 0\bmod p$ there is an integer
$0\leq r(\ell)\leq 4$ such that
 $$ 
5n(\ell) + \ell~=~r(\ell) p~.$$
A little thought reveals that, for $p\neq 5$, $r(\ell)$ has the property that it is the smallest positive
residue of \hbox{$b\ell$ mod 5} 
$$ 
r(\ell)=b\ell - 5\left[{b\ell\over 5}\right]~.$$

Returning to \eqref{tildeUone} we see that
 $$
\youth  (0)\ket{\ell}~=~(-1)^\ell p^r\,\prod_{i=1}^5 \left(\sum_{k_i=0}^\infty c_{k_i p + n}\, p^{k_i}
\left(-\p X_0X_i^5\right)^{k_i}\right)\,\ket{r(\ell)}~.$$

Now for any state $\ket{\chi}$ we have the following identity that holds in cohomology
 $$
\death _i\ket{\chi}~=~\Big(D_i + 5\p X_0X_i^5 -\p\vph Q\Big)\ket{\chi}~=~0$$
or equivalently
 $$
\left(-\p X_0X_i^5\right)\ket{\chi}~=~\hbox{e}^{\p\vph Q}\, {D_i\over 5}\, \hbox{e}^{-\p\vph Q}\ket{\chi}~.\eqlabel{coho}$$
Where we defer setting $\vph=0$ for the present.
Now it follows from this last relation that
 $$
\left(-\p X_0X_i^5\right)^m\ket{\chi}~=~\hbox{e}^{\p\vph Q}\, \left({D_i\over 5}\right)_m 
\hbox{e}^{-\p\vph Q}\ket{\chi}$$
A word of caution is warranted here. Note that the right hand side of this last relation contains the Pochhammer symbol 
 $$
\left({D_i\over 5}\right)_m~=~\left({D_i\over 5}\right)\left({D_i\over 5}+1\right)\ldots
\left({D_i\over 5}+m-1\right)$$
{\sl not} the power $(D_i/5)^m$. This is so because we must unwrap the powers of $X$ from the outside in order to use \eqref{coho}. Thus we have
$$\eqalign{
\left(-\p X_0X_i^5\right)^m\,\ket{\Ps}~
&=~-\p X_0X_i^5\,\Big(\left(-\p X_0X_i^5\right)^{m-1}\ket{\Ps}\Big)\cr
&=~\hbox{e}^{\p\vph Q}\, {D_i\over 5}\, \hbox{e}^{-\p\vph Q}
\Big(\left(-\p X_0X_i^5\right)^{m-1}\ket{\Ps}\Big)\cr
&=~\left(-\p X_0X_i^5\right)^{m-1}
\hbox{e}^{\p\vph Q}\left({D_i\over 5}+m-1\right)\hbox{e}^{-\p\vph Q}\ket{\Ps}~.\cr }$$
To naively unwrap the power from the inside would be wrong since the equalities are true in cohomology, so 
$\ket{\Ps}=\ket{\Ph}$ means, of course, that $\ket{\Ps}-\ket{\Ph} = \death\ket{\ch}$ for some 
$\ket{\ch}$. The formalism has the awkward feature that we can have $\ket{\Ps}=\ket{\Ph}$, in this sense,
yet $f(X)\ket{\Ps}\neq f(X)\ket{\Ph}$, in general, since $f(X)\death\ket{\ch}$ is not necessarily a derivative.
Worse still, it follows that we can have $\ket{\Ps}=0$ but $f(X)\ket{\Ps}\neq 0$. We could avoid this awkwardness by restricting the allowed operators to be those that commute with $\death$ however that would leave us with a very small class of operators; in the present case we would be limited to just the constants. An essential aim in the following is to study how the cohomology varies with $\vph$. It is precisely because the operators we use do not commute with $\death$ that they can map between spaces corresponding to different values of $\vph$. It is better therefore to proceed carefully with these caveats in mind.

Returning to our calculation, we take note of a relation established recently by Robert~
\Ref{\Robert}{A.~M.~Robert, ``The Gross-Koblitz Formula Revisited'',\\ Rend.~Sem.~Mat.~Univ.~Padova,
{\bf 106} 2001.}
 $$
\G_p(pz-a)~=~\sum_{k=0}^\infty c_{kp+a}\, p^k (z)_k~~;~0\leq a\leq p-1~~,~~z\in \Z_p$$
and we see that our expressions simplify considerably yielding
 $$
\youth  (0)\ket{\ell}~=~(-1)^\ell p^r\,\G_p^5\left({\ell\over 5}\right)\ket{r(\ell)}$$
where in writing this last relation we have used the fact that ${pr\over 5}-n={\ell\over 5}$.
The matrix $\bra{k}\youth  (0)\ket{\ell}$ is of size $5{\times}5$, has one entry in each row and
column and $\bra{0}\youth  (0)\ket{0}=1$. The characteristic equation of the $4{\times} 4$
block with $1\leq k,l\leq 4$, after the replacement $t\to t/p$ yields the the factor $R_{\bf 1}(t,0)$
of the zeta function of \cite{\CdORVII}. 

Recall that $r(\ell)$ is the reduction of $b\ell$ mod 5. We see from the following table that $b^\r\equiv 1$
mod 5 precisely when $p^\r\equiv 1$ mod 5, in fact, for $p\neq 2,5$, we have that $b$ is the smallest positive integer that is  $1/p \bmod 5$. For $\r=1$, that is $5|p-1$, we have $b=1$ and $r(\ell)=\ell$ so
the matrix is diagonal and one easily verifies the corresponding quartics $R_{\bf 1}(t,0)$ given in 
\cite{\CdORVII}. For
$\r=2$ we have $\youth  (0)^2\ket{\ell}=-p^5\ket{\ell},~1\leq \ell\leq 4$, leading to 
$R_{\bf 1}(t,0)=(1+p^3t^2)^2$ in these cases. For $\r=4$ the states form a cycle of length four and
$\youth  (0)^4\ket{\ell}=-p^{10}\ket{\ell},~1\leq \ell\leq 4$ leading to 
$R_{\bf 1}(t,0)=1+p^6t^4$.
\vskip-5pt
$$\vbox{\def\skip{\hskip10pt}
\offinterlineskip\halign{
\vrule\strut\hfil\hskip5pt #\hskip3pt\hfil\vrule height 13pt depth 6pt
&\hfil\skip $#$\skip\hfil\vrule&\hfil\skip$#$\skip\vrule&\hfil\skip$#$\skip\vrule&\hfil\skip$#$\skip\vrule\cr
\noalign{\hrule}
$p$ mod 5 & a& b& \hbox{relation}& \r\cr
\noalign{\hrule\vskip3pt\hrule}
1&{p-1\over 5} & 1 & b=1   & 1 \cr\noalign{\hrule}
2&{3p-1\over 5}& 3 & b^4\equiv 1 & 4 \cr\noalign{\hrule}
3&{2p-1\over 5}& 2 & b^4\equiv 1 & 4 \cr\noalign{\hrule}
4&{4p-1\over 5}& 4 & b^2\equiv 1 & 2 \cr\noalign{\hrule}
}}
$$
\vskip5pt
For $\vph=0$ it is straightforward to see that a basis of solutions to the equations 
$\death^*_\a \Ph^*(X)=0$ is provided by the dual states
 $$
\bra{\bf u}~=~X^{-{\bf u}}\prod_{i=1}^5\sum_{r_i=0}^\infty 
{\left({u_i\over 5}\right)_{r_i}\over (-\p X_0 X_i^5)^{r_i}}~~;~~0\leq u_i\leq 4~~,~~ 5|\deg({\bf u})$$
and that these states are dual to the states $X^{\bf v}$.
\subsection{Variation of structure}
We turn now to the case $\vph\neq 0$ and the operator
 $$\eqalign{
\youth(\vph)~&=~\aith _p \exp\Bigl(\p\bigl(W(X)-W(X)^p\bigr)\Bigr)\cr
~&=~\aith _p  \prod_{i=1}^5 \faith(X_0X_i^5)\, \faith(-\vph Q)\cr
~&=~\hbox{e}^{ \p \vph^p Q}\,\youth(0)\, \hbox{e}^{-\p\vph Q}~.\cr}$$
In order to evaluate the matrix for the operator $\youth(\vph)$ explicitly we
first take $\norm{\vph}_p<1$, for which the operators $\hbox{e}^{-\p\vph Q}$ and 
$\hbox{e}^{\p\vph^p Q}$ preserve $\ca{H}$, and we will later continue $\youth(\vph)$ to 
$\norm{\vph}_p=1$. 
In our discussion of cohomology in the previous section it was implicit that the parameters of $W$
satisfied relations analogous to $\vph^p=\vph$. Now we have $\norm{\vph}_p<1$ so a term $\vph^p Q^p$ arises in $W(X)^p$ and this leads to the presence of $\vph^p$ in the expression for $\youth(\vph)$.

We wish to study how the matrix $\youth  (\vph)$ acts on the cohomology groups
$\ca{H}/\death$. To emphasize the dependence of the states in the quotient space on
the parameter we append a $\vph$ to the states. Thus the state $\ket{{\bf v}}_\vph$, for
example, denotes the equivalence class of $X^{\bf v}$ in $\ca{H}/\death$. The parameter
dependence arises because the states now correspond to equivalence classes and the
$\death$-operator depends on $\vph$. We write $\death(\vph)$ to emphasize this dependence; 
setting $W = W_0 - \vph Q$ we have
 $$
\death (\vph) = \hbox{e}^{\p\vph Q} \hbox{e}^{-\p W_0} D\, \hbox{e}^{\p W_0} \hbox{e}^{-\p\vph Q} =\hbox{e}^{\p\vph Q}\,\death (0)\, \hbox{e}^{-\p\vph Q}~.\eqlabel{Drels}$$

Now, by writing out the derivative explicitly, we see that
 $$
{1\over 5}\death_i(\vph)\ket{\Ps}~=~\left(\p X_0X_i^5 + \hbox{e}^{\p\vph Q} 
{D_i\over 5} \hbox{e}^{-\p\vph Q}\right)\ket{\Ps}$$  

As a relation on $\ca{H}/\death (\vph)$ this becomes
 $$
\left(-\p X_0X_i^5\right)\ket{\Ps}_\vph ~=~ 
\hbox{e}^{\p\vph Q} {D_i\over 5} \hbox{e}^{-\p\vph Q}\ket{\Ps}_\vph~.$$ 
The corresponding relation on $\ca{H}/\death (0)$ is
 $$
\left(-\p X_0X_i^5\right)\ket{\Ps}_0~=~{D_i\over 5} \ket{\Ps}_0~.\eqlabel{cohozero}$$

In virtue of \eqref{Drels}, we also have 
$\hbox{e}^{-\p\vph Q}\death (\vph) = \death (0)\hbox{e}^{-\p\vph Q}$. It
follows that
 $$
\hbox{e}^{-\p\vph Q}\Bigl( \ket{\Ps}+\death (\vph)\ket{\ch}\Bigr)~=~
\ket{\hbox{e}^{-\p\vph Q}\,\Ps} + \death (0)\ket{\hbox{e}^{-\p\vph Q}\,\ch}~.$$
From this and an analogous relation involving $\hbox{e}^{\p\vph Q}$ we learn that 
 $$
\hbox{e}^{-\p\vph Q}:\, \ca{H}/\death(\vph) \to \ca{H}/\death(0)~~~\hbox{and}~~~
\hbox{e}^{\p\vph Q}:\, \ca{H}/\death(0) \to \ca{H}/\death(\vph)~.$$
A restatement of the above is that
$$
\hbox{e}^{-\p\vph Q}\ket{\Ps}_\vph~=~\ket{\hbox{e}^{-\p\vph Q}\Ps}_0~~~\hbox{and}~~~
\hbox{e}^{\p\vph Q}\ket{\Phi}_0~=~\ket{\hbox{e}^{\p\vph Q}\Phi}_\vph~.
$$
We find it convenient to build the exponential factors into the notation and write
$$
\ket{\Ps;\, \vph}~\define~\hbox{e}^{-\p\vph Q}\ket{\Ps}_\vph~~~\hbox{and}~~~
\ket{\Ps;0}~\define~\ket{\Ps}_0~.
$$

We apply these considerations to the operator $\youth  (\vph)$ which is a composition of maps between the following spaces
 $$
\matrix{\ca{H}/\death (\vph)& \buildrel{\hbox{\sevenrm e}^{-\p\vph Q}}\over\arr    &\ca{H}/\death(0)\cropen{5pt}
{\scriptstyle\sevenyouth(\vph)}\downarrow~~~&&~~~\downarrow{\scriptstyle\sevenyouth(0)}\cropen{5pt}
\ca{H}/\death(\vph^p)& \buildrel{\hbox{\sevenrm e}^{\p\vph^p Q}}\over\longleftarrow &\ca{H}/\death(0)\cr}
$$
We will understand the matrix for $\youth(\vph)$ to have components
 $$
U_{j\ell}(\vph) ~=~ {}_{\vph^p}\!\bra{j}\hbox{e}^{\p \vph^p Q}\,\youth  (0)\,
\hbox{e}^{-\p \vph Q}\ket{\ell}_\vph
~=~\bra{j;\,\vph^p}\youth(0)\ket{\ell;\,\vph}~.
$$
The operator $\youth  (\vph)$ maps between spaces that are in general different but which become the same
when $\vph^p=\vph$.  Now for the state on the right of this last matrix element we can write
 $$\eqalign{
\ket{\ell;\,\vph}~&=~\ket{\hbox{e}^{-\p \vph Q}(-\p Q)^\ell;\,0}\cr
&=~\sum_{n=0}^\infty{\vph^{n-\ell}\over (n-\ell)!}\,\ket{n;\, 0}\cr 
&=~\sum_{j=0}^4\sum_{m=0}^\infty {\vph^{5m+j-\ell}\over
(5m+j-\ell)!}\,\ket{5m+j;\,0}\cr  
&=~\sum_{j=0}^4\sum_{m=0}^\infty {\vph^{5m+j-\ell}\over (5m+j-\ell)!}\,
\left({j\over 5}\right)_m^5\,\ket{j;\,0}\cr}$$
where the third line follows from breaking up the $n$-sum by writing $n=5m+j$, $0\leq j\leq 4$, and 
the fourth by noting that
 $$
(-\p Q)^{5m}\,\ket{\Ps;\,0}~=~\prod_{i=1}^5 \left(-\p X_0X_i^5\right)^m\,\ket{\Ps;\,0}
~=~ \prod_{i=1}^5\left({D_i\over 5}\right)_m\ket{\Ps;\,0}~.\eqlabel{Qpowers}$$
In this way we see that
 $$
\ket{\ell;\,\vph}~=~\sum_{j=0}^4\, \ket{j;\,0}E_{j\ell}(\vph)\eqlabel{Qleqn}$$
with
 $$
E_{j\ell}(\vph)~=~\sum_{m=0}^\infty {\vph^{5m+j-\ell}\over (5m+j-\ell)!}\,
\left({j\over 5}\right)_m^5~.\eqlabel{Qleqntwo}$$
It follows also that
 $$
\bra{j;\,\vph^p}~=~
\bra{(-\p Q)^j \hbox{e}^{-\p \vph^p Q};\,0\,}~=~\sum_{k=0}^4\, E_{jk}^{-1}(\vph^p) \bra{k;\,0}$$
and hence we arrive at the expression
 $$
U(\vph)~=~E^{-1}(\vph^p)\, U(0)\, E(\vph)~.\eqlabel{Uphi}$$

We are seeking to calculate the $5{\times}5$ determinant $\det(1-U(\vph)T/p)$. The matrices $E^{-1}$, $U(0)$
and $E$ and hence also $U(\vph)$ all have a first row $(1,0,0,0,0)$. Thus apart from a factor of $(1-T/p)$
the $5{\times}5$ determinant reduces to a $4{\times}4$ determinant of the same form but with the matrix
indices running over the range $1\leq j,\ell\leq 4$. We shall abuse notation by denoting these reduced
matrices by the same symbols as previously. 

The matrix $E_{j\ell},\,1\leq j,\ell \leq 4$ is a Wronskian matrix as we see by noting that
$$
E_{j\ell}(\vph)=\left({d\over d\vph}\right)^{\ell-1}\,E_{j1}(\vph)$$

and that $\vph E_{j1}$ satisfies the Picard-Fuchs\Footnote{The Picard-Fuchs operator that we give here is related to the corresponding Picard-Fuchs operator, $\ca{L}$, of \cite{\CdORVI} by 
$\ca{L}_\vph={\vph^5\over 5}\ca{L}$.} equation
$$
\ca{L}_\vph\Big(\vph E_{j1}(\vph)\Big) =0~,~~~\hbox{with}~~~
\ca{L}_\vph=\left({\vph\over 5}\right)^5 \d_\vph^4 - \prod_{i=1}^4\left(\d_\vph - i\right) $$
with $\d_\vph=\vph {d\over d\vph}$. Being a Wronskian the determinant of $E$ has a simple form
$$
\det E={1\over \left(1 - \left({\vph\over 5}\right)^5\right)^2}~.$$
The components of the matrices $E(\vph)$ and $E^{-1}(\vph)$ are power series in $\vph$ with coefficients that
are p-adic fractions. However it appears to be the case, on the basis of numerical experiment, that
$U(\vph)$ is a matrix whose components are power series with coefficients that are p-adic integers.
Integrality of these coefficients would follow it one could show that the basis we use `comes from' crystalline cohomology. One may well be able to do this using results from \SS3 of~
\Ref{\AbbotKedlayaRoe}{T. G. Abbott, K. S. Kedlaya and D. Roe,``Bounding Picard numbers of surfaces using p-adic cohomology'',  to appear in Arithmetic, Geometry and Coding Theory, (AGCT 2005), 
Societ\'{e} Math\'{e}matique de France.}. 
In any event equation
\eqref{Uphi} provides a power series expansion for $U(\vph)$. This series converges for $\norm{\vph}_p<1$ but not
for $\norm{\vph}_p = 1$ since the coefficients do not tend to zero. It~has, however, been shown by Lauder~
\Ref{\Lauder}{A. Lauder, ``Counting solutions to equations in many variables over finite fields'', Foundations of Computational Mathematics {\bf 4} No. 3 pp221-267 2004.}\
(see especially the Appendix) that the offending coefficients can however be
summed to a rational function. For the present case, numerical experiment suggests that we may write
 $$
U\Bigl(\teich(5)\ps\Bigr)~=~\left({1-\ps^{5\hphantom{p^n}}\over 1 - \ps^{5p^n}}\right) 
\widetilde{U}_n\Big(\teich(5)\ps\Big) + \ord{p^N} \eqlabel{Utilde}$$
with $\widetilde{U}_n(\vph)$ a polynomial in $\vph$ of order approximately $5p^n$ and $N>n$. For example if
$p=3,7$ or 11 and $n=3$ then
$N=8$ (in fact for $p=3$ we have $N=9$). Thus we may compute $\widetilde{U}(\vph)$ to arbitrary p-adic
precision by taking $n$ sufficiently large and then setting $\ps^p=\ps$ in $\widetilde{U}_n$. It is a
consequence of the relation \eqref{Utilde} that the limit
 $$
\widetilde{U}(\vph) = \lim_{n\to\infty}\widetilde{U}_n(\vph)$$
exists and is a convergent series for $\norm{\vph}_p=1$. Note also that if we understand the rational function as a limit we have
 $$
{1-\ps^{5\hphantom{p^n}}\over 1 - \ps^{5p^n}}~=~
\cases{1~;&$\ps=\teich(u),~u\in\Fp,~u^5\neq1$\cropen{5pt}
         \displaystyle{1\over p^n}~;& $\ps^5=1~.$\cr}$$
We find that $p^n|\widetilde{U}_n(\vph)$ for $\ps^5=1$ so that we may use \eqref{Utilde} to compute $U$ even
for these values.
Take now $\vph=\teich(5u),~u\in\Fp$, with $u^5\neq1$ then we have
 $$
\det\left(1-\widetilde{U}(\vph)\,T/p\right)~=~
1+\tilde{a}(\vph)\,T+\tilde{b}(\vph)\,pT^2+\tilde{a}(\vph)\,p^3T^3+p^6T^4$$
with the coefficients most easily calculated as
 $$
\tilde{a}~=~-{1\over p}\,\tr\,\widetilde{U}~~~ \hbox{and}~~~ 
\tilde{b}~=~{1\over 2p^3}\left( \big(\tr\,\widetilde{U}\big)^2 - \tr\big(\widetilde{U}^2\big)\right)~.$$
In this way we recover the values of the coefficients given in the tables of \cite{\CdORVII}. 
\subsection{The other monomials}
By applying a procedure closely analogous to that of \SS3.1 we may calculate the matrix $U$ in a basis
corresponding to the monomials $\ket{\bf v}=(-\p)^{v_0} X^{\bf v}$. We find that
 $$
\youth  (0)\ket{\bf v}~=~(-1)^{v_0} p^{r_0} \prod_{i=1}^5 \G_p\left({v_i\over 5}\right)\, \ket{\bf r}
\eqlabel{Uzerov}$$
where $r_i$ is the smallest positive residue of $bv_i\bmod5$ and $b,\,1\leq b\leq 4$, is again such that
$bp=1\bmod 5$.
It is straightforward also to find the relations analogous to \eqref{Qleqn} and \eqref{Qleqntwo}. Let
$X^{\tilde{\bf v}}$ be the monomial resulting from the extraction of as many powers of $Q$ from $X^{\bf v}$
as possible, so that $X^{\bf v}=Q^\ell X^{\tilde{\bf v}}$ and at least one component of $\tilde{\bf v}$ is
zero.
 $$
\ket{\hbox{e}^{-\p \vph Q}(-\p)^{v_0}X^{\bf v};\,0}~=~
\ket{\hbox{e}^{-\p \vph Q}(-\p)^{\ell+\tilde{v}_0} Q^\ell X^{\tilde{\bf v}};\,0}~=~
\sum_{j=0}^4\, \ket{j\bone+\tilde{\bf v};0}\,E_{\tilde{\bf v},j,l}(\vph) \eqlabel{Xveqn}$$
with
 $$
E_{\tilde{\bf v},j,l}(\vph)~=~\sum_{m=0}^\infty {\vph^{5m+j-\ell}\over
(5m+j-\ell)!}\,\prod_{i=1}^5\left({j+\tilde{v}_i\over 5}\right)_m~.\eqlabel{Xveqntwo}$$

To apply this result consider the family of monomials obtained from a given monomial $X^{\bf v}$ by multiplying successively by powers of $Q$ and then reducing the exponents mod 5. The families that descend in this way from our representative quintics are displayed in the~table:
 $$
\vbox{\def\skip{\hskip10pt}
\offinterlineskip\halign{
\vrule\strut\hfil\skip $#$\skip\vrule height 13pt depth 7pt
&\hfil\skip $#$\skip\vrule&\hfil\skip$#$\skip\vrule&\hfil\skip$#$\skip\vrule\cr
\noalign{\hrule}
(1,4,0,0,0)&(3,2,0,0,0) &(3,1,1,0,0)&(2,2,1,0,0)\cr
\noalign{\hrule\vskip3pt\hrule}
(1,4,0,0,0)&(3,2,0,0,0) &(3,1,1,0,0)&(2,2,1,0,0)\cr
(2,0,1,1,1)&*\,(4,3,1,1,1)&*\,(4,2,2,1,1)&*\,(3,3,2,1,1)\cr
*\,(3,1,2,2,2)&(0,4,2,2,2)&(0,3,3,2,2)&*\,(4,4,3,2,2)\cr
*\,(4,2,3,3,3)&(1,0,3,3,3)&*\,(1,4,4,3,3)&(0,0,4,3,3)\cr
(0,3,4,4,4)&*\,(2,1,4,4,4)&(2,0,0,4,4)&(1,1,0,4,4)\cr
\noalign{\hrule\vskip3pt\hrule}
(a,b)=({2\over 5},{2\over 5})&(a,b)=({1\over 5},{1\over 5})&
(a,b)=({1\over 5},{2\over 5})&(a,b)=({1\over 5},{3\over 5})\cr
\noalign{\hrule}
}}
$$
In each column of the Table there are two monomials that are distinguished by a $*$. These are monomials that have $\ell\neq0$. Now it is clear that the relation \eqref{Xveqn} acts within each family and a little thought reveals that, on a family, the operator $\hbox{e}^{-\p\vph Q}$ acts as a $5{\times}5$ matrix which, relative to a basis in which the three monomials with $\ell=0$ are taken first and the two with $\ell\neq 0$ are taken last, has the block structure
$$
\pmatrix{\bf 1&S\cr \bf 0 & G\cr}$$
where the $\bf 1$ denotes a $3{\times}3$ matrix and $G$ denotes a $2{\times}2$ matrix. The unit matrix arises from the $j=0$ terms in \eqref{Xveqn} when $\ell=0$. Now the inverse of such a block matrix has a similar form, in fact
$$
\pmatrix{\bf 1&S\cr \bf 0 & G\cr}^{-1}~=~\pmatrix{\bf 1&-SG^{-1}\cr \bf 0 & G^{-1}\cr}~.$$
The operator $\youth  (0)$ will in general act between monomials from different families but the operator preserves the number of zeros of a vector of exponents. Thus the vectors that have $\ell\neq 0$ are mapped among themselves. By considering all the vectors together we can form big matrices for the operators       
$\hbox{e}^{-\p\vph Q}$ and $\hbox{e}^{\p\vph^p Q}$, but provided we write a basis with the vectors that have  $\ell=0$ before those with $\ell\neq 0$ these big matrices will still take a block triangular form of the type we are discussing. In this basis the operator $\youth  (0)$ will be block-diagonal. It follows that the matrix                 
$({\bf 1}-\youth  (\vph)T/p)$ is block upper triangular and hence its determinant reduces (apart from a term 
$(1-T/p)^n$, that is independent of $\vph$) to the determinant of a matrix evaluated on the monomials that have 
$\ell\neq 0$.

If we evaluate the operator $\hbox{e}^{-\p\vph Q}$ with respect to the bases provided by these vectors with              
$\ell\neq 0$ we find, on using \eqref{Xveqn} and the multiplication formula for the $\G$-function in the form 
$$
(5m)!~=~5^{5m}\, m!\, \left({1\over 5}\right)_{\! m} \left({2\over 5}\right)_{\! m} 
\left({3\over 5}\right)_{\! m} \left({4\over 5}\right)_{\! m}~,$$
that it corresponds to a matrix that is essentially a Wronskian matrix of hypergeometric functions. Setting 
$z=\left({\vph\over 5}\right)^5$,
$$
f(a,b;z)={}_2F_1(a,b;a+b;z)~~~\hbox{and}~~~
g(a,b;z)={z^{1-a-b}\over 1-a-b}\, {}_2F_1(1-a,1-b;2-a-b;z)~,$$
we have
$$
G(a,b;z)~=~\pmatrix{f(a,b;z)& z^{a+b}{d\over dz}\, f(a,b;z)\cropen{10pt}
g(a,b;z)&z^{a+b}{d\over dz}\, g(a,b;z)\cr}$$
where, for each family, the parameters $a$ $b$ take the values shown in the Table. In each case we have the relation
$$
\det G(a,b;z)={1\over 1-z}~.$$
The values of the $a$ and $b$ coefficients given in the Table differ from those that correspond to the $\ca{A}$ and 
$\ca{B}$ curves of \cite{\CdORVII} but this is due merely to the fact that here we have hypergeometric functions with argument $z$ while in \cite{\CdORVII} we were dealing with functions of argument $1/z$. The functions
$$
z^{-a}{}_2F_1\left(a,1-b;1+a-b;{1\over z}\right) ~~~\hbox{and}~~~
z^{-b}{}_2F_1\left(1-a,b;1-a+b;{1\over z}\right)
$$
satisfy the same differential equation as ${}_2F_1(a,b;a+b;z)$.

The block structure of $\youth  (0)$ depends on the value of $\r$. We will illustrate the structure for $\r=4$, since the other cases are simpler. 
The matrix $\youth  (0)$ relates, for the case $\r=4$, a vector $\bf v$ with the mod 5 reduction of $2\bf v$. This process interchanges the permutations of the $(4,1,0,0,0)$ family with the $(3,2,0,0,0)$ family. We take as a basis the monomials:
$$\eqalign{
&(1,3,2,2,2)\cropen{-2pt}
&(2,4,3,3,3)\cropen{3pt}
&(2,1,4,4,4)\cropen{-2pt}
&(4,3,1,1,1)\cropen{3pt}
&(4,2,3,3,3)\cropen{-2pt}
&(3,1,2,2,2)\cropen{3pt}
&(3,4,1,1,1)\cropen{-2pt}
&(1,2,4,4,4)~.\cr}$$
These are taken in pairs, alternately, from the $(4,1,0,0,0)$ and $(3,2,0,0,0)$ families and their permutations. With respect to this basis the matrix $\youth  (\vph)$ takes the form
$$
\pmatrix{\widetilde{G}&0&0&0\cr  0&\widetilde{H}&0&0\cr 0&0&\widetilde{G}&0\cr
 0&0&0&\widetilde{H}}
 \pmatrix{0&0&0&V\cr W&0&0&0\cr 0&V&0&0\cr 0&0&W&0\cr}
 \pmatrix{G&0&0&0\cr  0&H&0&0\cr 0&0&G&0\cr 0&0&0&H\cr}
 =\pmatrix{0&0&0&Y\cr Z&0&0&0\cr 0&Y&0&0\cr 0&0&Z&0\cr}$$
 where the blocks correspond to $2{\times}2$ matrices, $Y=\widetilde{G}VH$ and $Z=\widetilde{H}WG$.
 Thus the determinant we require takes the form
 $$
 \det\left({\bf 1}-\youth\, {T\over p}\right) =
  \det\pmatrix{\bf 1&0&0&-TY/p\cr -TZ/p&\bf 1&0&0\cr 0&-TY/p&\bf 1&0\cr 
 0&0&-TZ/p&\bf 1\cr} = \det\left({\bf 1}-{T^4\over p^4}(YZ)^2\right)$$
 The last equality follows on factoring the matrix into the product of an upper triangular and lower triangular matrix
 $$\eqalign{
&\pmatrix{\bf 1&0&0&-TY/p\cr -TZ/p&\bf 1&0&0\cr 0&-TY/p&\bf 1&0\cr 
0&0&-TZ/p&\bf 1\cr} =\cropen{10pt}
&\hskip0in
\pmatrix{\bf 1&-T^3 YZY/p^3&-T^2 YZ/p^2&-T Y/p\cr 0&\bf 1&0&0\cr 0&0&\bf 1&0\cr 0&0&0&\bf 1\cr}\!
\pmatrix{{\bf 1}-T^4 (YZ)^2/p^4&0&0&0\cr  -T Z/p&\bf 1&0&0\cr 0&-T Y/p&\bf 1&0\cr 0&0&-T Z/p&\bf 1\cr}
\cr}$$
It follows from these considerations that 
$$ 
\det\left({\bf 1}-\youth\,  {T\over p}\right) = 1 + p^6 c(\vph)\,T^4 + p^{12}T^8~.$$
The explicit factors of $p$ follow from counting the explicit powers that appear in \eqref{Uzerov}. 
The coefficient $c(\vph)$ has, for $\vph=\teich(5)\ps$, a structure similar to that which we have seen in the previous section
$$
c(\vph) = {1\over p^4}\Tr\Big((YZ)^2\Big)=
\left({1-\ps^{5\hphantom{p^n}}\over 1 - \ps^{5p^n}}\right) 
\tilde{c}_n\Big(\teich(5)\ps\Big) + \ord{p^N}$$
which again allows us to recover the results of \cite{\CdORVII}.
\newpage
\section{open}{Open Problems}
We list here three open problems related to the present work:
\vskip-20pt
\subsection{Special Geometry}
A formulation of $p$-adic special geometry valid throughout the moduli space would seem to be a necessary prerequisite for a proper discussion of the arithmetic properties of conifolds and attractor geometries.
\vskip-10pt
\subsection{The $\vph=\infty$ limit}
Related to the above is the difficulty of developing a formalism analogous to Dwork analysis presented here but with the ability to expand the operator $\youth(\vph)$ about $\vph=\infty$ instead of expanding, as here, about $\vph=0$. The manifold corresponding to $\vph=\infty$ is highly singular and this has, so far, prevented the application of the methods illustrated here. A specific question is what is the `correct' form for the $\zeta$-function for the quintic for $\vph=\infty$.
\vskip-10pt
\subsection{Truncated periods vs. infinite series}
In this work we have calculated the numbers of points of the manifold, say $N_1(\vph)$, in terms of the periods, with the periods given by infinite series. In \cite{\CdORVI} these same numbers are calculated in terms of {\sl truncated periods}, that is series truncated to the first $p$ terms whose argument is the parameter evaluated on the Teichm\"{u}ller points. Since, in the latter approach, one deals with finite series there is no need for the process of analytic continuation that has occupied us here. It is of interest to understand how these two calculations are related.
\vskip1in
\leftline{\bf Acknowledgements}
It is a pleasure to acknowledge fruitful conversations with Fernando Rodr\'{\i}guez-Villegas and Duco Van Straten and to thank Alan Lauder for much patient instruction.
\newpage
\frenchspacing
\immediate\closeout\referencewrite\referenceopenfalse
\line{\fourteenbold\hfil References\hfil}\bigskip\parindent=0pt\input referenc.texauxil
\bye